%% file: main.tex
\RequirePackage{fix-cm}
\documentclass[pdftex,twocolumn,final,epjc3]{svjour3}  

\usepackage{xspace}
\usepackage{enumitem}
\usepackage{amsmath}
\usepackage{amssymb}
\usepackage{nicefrac}
\usepackage{xcolor}
\usepackage{graphicx}
\usepackage{hyperref}
\usepackage{cite}
\usepackage{wasysym}

\usepackage{cuted}
\usepackage{etoolbox}
\AtBeginEnvironment{strip}{\leavevmode}

\smartqed
\journalname{Eur. Phys. J. C}

\input{definitions}
\input{comments}

\begin{document}

\title{Artificial neural network modelling of generalised parton distributions}

\author{
H.~Dutrieux \thanksref{emailHD,addressCEA} 
\and
O.~Grocholski\thanksref{emailOG,addressUW}
\and
H.~Moutarde\thanksref{emailHM,addressCEA}
\and
P.~Sznajder\thanksref{emailPS,addressNCBJ}
}

\thankstext{emailHD}{e-mail: herve.dutrieux@cea.fr}
\thankstext{emailOG}{e-mail: oskar.grocholski@desy.de}
\thankstext{emailPS}{e-mail: pawel.sznajder@ncbj.gov.pl}
\thankstext{emailHM}{e-mail: herve.moutarde@cea.fr}

\institute{
IRFU, CEA, Universit\'e Paris-Saclay, F-91191 Gif-sur-Yvette, France \label{addressCEA}
\and
~Deutsches Elektronen-Synchrotron DESY, Notkestr.
85, 22607 Hamburg, Germany  \label{addressUW}
\and
~National Centre for Nuclear Research (NCBJ), Pasteura 7, 02-093 Warsaw, Poland \label{addressNCBJ}
}

\date{Received: date / Accepted: date}

\maketitle

\sloppy


\input{sec_abstract}
\input{sec_introduction}
\input{sec_theory}
\input{sec_xxi}
\input{sec_ba}
\input{sec_ba_fits}
\clearpage
\input{sec_summary}
\input{sec_ack}

\bibliographystyle{spphys}
\bibliography{bibliography}

\end{document}

%% file: definitions.tex
\newcommand{\muF}{\mu_{\mathrm{F}}}
\newcommand{\muFRef}{\mu_{\mathrm{F}, 0}}

\newcommand{\eg}{\textit{e.g.}\xspace}
\newcommand{\ie}{\textit{i.e.}\xspace}

\newcommand{\etc}{\textit{etc.}\xspace}

\def\dx{\mathrm{d}x}
\def\mod{\mathrm{mod}}

%% file: comments.tex
\newcounter{comment}

{\refstepcounter{comment}%
\begin{quote}
\color{blue}{#1~ \normalsize \textbf{\underline{Comment} $\sharp$\thecomment~by~PS:~}}}%
{\end{quote}}

{\refstepcounter{comment}%
\begin{quote}
\color{violet}{#1~ \normalsize \textbf{\underline{Comment} $\sharp$\thecomment~by~HM:~}}}%
{\end{quote}}

{\refstepcounter{comment}%
\begin{quote}
\color{red}{#1~ \normalsize \textbf{\underline{Comment} $\sharp$\thecomment~by~OG:~}}}%
{\end{quote}}

{\refstepcounter{comment}%
\begin{quote}
\color{magenta}{#1~ \normalsize \textbf{\underline{Comment} $\sharp$\thecomment~by~HD:~}}}%
{\end{quote}}

%% file: sec_abstract.tex
\begin{abstract}
We discuss the use of machine learning techniques in  effectively nonparametric modelling of generalised parton distributions (GPDs) in view of their future extraction from experimental data. Current parameterisations of GPDs suffer from model dependency that lessens their impact on phenomenology and brings unknown systematics to the estimation of quantities like Mellin moments. The new strategy presented in this study allows to describe GPDs in a way fulfilling theory-driven constraints, keeping model dependency to a minimum. Getting a better grip on the control of systematic effects, our work will help the GPD phenomenology to achieve its maturity in the precision era commenced by the new generation of experiments.
\end{abstract}

%% file: sec_introduction.tex
\section{Introduction}
\label{sec:introduction}
 
Generalised parton distributions (GPDs) \cite{Mueller:1998fv, Ji:1996ek, Ji:1996nm, Radyushkin:1996ru, Radyushkin:1997ki} are widely recognised as one of the key objects to explore the structure of hadrons. They encompass information coming from one-dimensional parton distribution functions (PDFs) and elastic form factors (EFFs). GPDs allow for a hadron tomography \cite{Burkardt:2000za, Burkardt:2002hr, Burkardt:2004bv}, where densities of partons carrying a fraction of hadron momentum are studied in the plane perpendicular to the hadron's direction of motion. GPDs also provide access to the  matrix elements of the energy-momentum tensor \cite{Ji:1996ek, Ji:1996nm, Ji:1997gm}, making it possible to evaluate the total angular momentum and ``mechanical'' properties of hadrons, like pressure and shear stress at a given point of space \cite{Polyakov:2002wz, Polyakov:2002yz, Polyakov:2018zvc, Lorce:2018egm}. 

Experimental access to GPDs is mainly possible thanks to exclusive processes occurring on hadrons remaining coherent after a hard scale interaction. Some notable processes of this type are deeply virtual Compton scattering (DVCS) \cite{Ji:1996nm}, time-like Compton scattering (TCS) \cite{Berger:2001xd} and deeply virtual meson production (DVMP) \cite{Radyushkin:1996ru,Collins:1996fb}. All of them allow to study transitions of hadrons from one state to another, with a unique insight into changes taking place at the partonic level. GPDs have been primarily studied in leptoproduction experiments, in particular those conducted in JLab, DESY and CERN, and are key objects of interest in programmes of future electron-ion colliders, like EIC \cite{Accardi:2012qut, AbdulKhalek:2021gbh}, EicC \cite{Anderle:2021wcy} and LHeC \cite{AbelleiraFernandez:2012cc}. Many data sets have already been collected and reviewed for DVCS \cite{Kumericki:2016ehc,dHose:2016mda} and DVMP \cite{Favart:2015umi}, while the first measurement for TCS has been completed recently \cite{CLAS:2021lky}.

Although several datasets are already available for fits, the extraction of GPDs is far from being satisfactory. The main reasons are:
\begin{enumerate}[label=\emph{\roman*})]
	\item sparsity of available information. GPDs are multidimensional functions, so one needs much more data to constrain them in the phase-space of kinematic variables, comparing to \eg one-dimensional PDFs.  Furthermore, one needs to cover this phase-space by data collected for various processes and experimental setups, which is required to distinguish between many types of GPDs and contributions coming from various quark flavours and gluons.  
	\item complexity of extraction. In order to fully benefit from available sources of information about GPDs, such as data collected in exclusive measurements, PDFs and EFFs for boundary conditions, lattice QCD,  one needs to know and implement links connecting those sources with GPDs. The extraction of GPDs requires a deconvolution of amplitudes measured in exclusive processes, which is not trivial and in some cases does not even possess a unique solution \cite{Bertone:2021yyz}. Both the evolution of GPDs and the description of exclusive processes must be understood to a level allowing for a robust extraction. This requires a substantial effort and a careful design of tools aggregating GPD-related theory developments. 
	\item model dependency. Currently available phenomenological GPD models, like GK \cite{Goloskokov:2005sd, Goloskokov:2007nt, Goloskokov:2009ia} and VGG \cite{Vanderhaeghen:1998uc, Vanderhaeghen:1999xj, Goeke:2001tz, Guidal:2004nd}, use similar Ans{\"a}tze with a rigid form and therefore cannot be considered as diverse sources for the estimation of model uncertainty. The severity of such a model dependency and its impact on \eg the extraction of orbital angular momentum have never been studied in a systematic way, partially because there were no tools to do so.
\end{enumerate}
The answer of the GPD community to those problems so far has been as follows. More data sensitive to GPDs will be delivered by a next generation of experiments. These experiments, performed in many laboratories and with multiple setups, will provide the much needed input to the GPD phenomenology. We witness a substantial progress in the description of exclusive processes and in the systematic use of GPD evolution equations \cite{Bertone:2021}. The development of lattice QCD techniques is accelerating, as proved by Ref.~\cite{Constantinou:2020pek}. The PARTONS \cite{Berthou:2015oaw} and GeParD \cite{Kumericki:GeParD} projects provide aggregation points for GPD-related developments and practical know-how in phenomenology methods. Techniques to stress model dependency are developed for the extraction of amplitudes, the latter used to access the ``mechanical'' properties of nucleons \cite{Kumericki:2019ddg, Dutrieux:2021nlz}. However, to the best of our knowledge, no substantial progress has been recently made to address the problem of model dependency of GPDs extractions.

In this Article, we discuss new ways of modelling GPDs which provide flexible parameterisations inspired by artificial neural network (ANN) techniques. These parameterisations may be used as tools to study the model dependency affecting extractions of GPDs and derived quantities, like orbital angular momentum of hadrons. We address aspects of a practical nature, \eg dealing with many free parameters. Our work aims in providing a better grip on the control of systematic effects, much needed in front of the precision era of GPD extractions. 

The Article is organised as follows. In Sect. \ref{sec:theory} we provide the theoretical background for our line of research. The modelling of GPDs in $(x,\xi)$-space is given in Sect.~\ref{sec:xxispace}, while that in $(\beta, \alpha)$-space is discussed in Sect.~\ref{sec:baspace}. We provide the summary in Sect.~\ref{sec:summary}.

%% file: sec_theory.tex
\section{Theoretical background}
\label{sec:theory}

GPDs are functions of three variables: the average longitudinal momentum fraction carried by the active parton, $x$, the skewness $\xi$, which describes the longitudinal momentum transfer, and the four-momentum transfer to the hadron target, $t$. GPDs also depend on the factorisation scale, $\muF$, being the variable entering evolution equations. A knowledge of these equations allows to run the evolution starting from a reference scale, $\muFRef$, where GPD models are defined, to any $\muF$. We therefore only consider GPDs at $\muFRef$, and in the following we suppress the $\muF$ dependence for brevity. For illustration we will only consider the GPD $H(x, \xi, t)$ for a quark of unspecified flavour, but we note that the discussion can be easily extended for other GPDs. 
$H(x, \xi, t)$ is a real and $\xi$-even function as a consequence of QCD invariance under discrete symmetries. Without loss of generality, we only consider positive $\xi$ in this work.

GPD models must fulfil a set of theory driven constraints:
\begin{enumerate}[label=\emph{\roman*})]
\item reduction to the PDF $q(x)$ in the forward limit:
	\begin{equation}
		H(x,\xi = 0, t = 0) \equiv q(x) \,,
		\label{eq:theory:pdf}
	\end{equation}
\item polynomiality \cite{Ji:1998pc,Radyushkin:1998bz,Polyakov:1999gs}, which is required to keep GPDs invariant under Lorentz transformation. The property states that any Mellin moment of a GPD, $\mathcal{A}_n$, is a fixed-order polynomial in $\xi$: 
	\begin{flalign}
		\mathcal{A}_{n}(\xi, t) & = 
		\int_{-1}^{1} \dx x^n H(x,\xi, t) \nonumber \\
		& = \sum_{\substack{j=0 \\ \mathrm{even}}}^{n} \xi^{j} A_{n, j}(t) + \mod(n, 2) \xi^{n+1} A_{n, n+1}(t) \,,
		\label{eq:theory:polynomiality}
	\end{flalign}
	where we call $A_{n,j}(t)$ a Mellin coefficient. We note that $\mathcal{A}_{0}(\xi, t) = A_{0, 0}(t) \equiv F_{1}(t)$ is the Dirac EFF form factor. The polynomiality ``entangles'' the $x$ and $\xi$ dependencies.  
\item positivity constraints \cite{Radyushkin:1998es, Pire:1998nw, Diehl:2000xz, Pobylitsa:2001nt, Pobylitsa:2002gw, Pobylitsa:2002iu, Pobylitsa:2002vi, Pobylitsa:2002vw, Pobylitsa:2002ru}, which are inequalities ensuring positive norms in the Hilbert space of states. Because of the variety of these inequalities, we do not quote them all here. Instead, we refer the Reader to the aforementioned references, and only note that the constraints typically involve several types of GPDs. In the following we will illustrate how to deal with the positivity constraints with this simple inequality \cite{Radyushkin:1998es, Pire:1998nw},
	\begin{equation}
	|H(x,\xi,t)| \leq \sqrt{q\left(\frac{x+\xi}{1+\xi}\right)q\left(\frac{x-\xi}{1-\xi}\right)\frac{1}{1-\xi^2}} \,,
	\label{eq:theory:positivity}
	\end{equation}
	which is applicable in the DGLAP region, \ie for $x>\xi$.
\end{enumerate}
For the sake of completeness, we write a direct relation between Mellin moments, which are extensively discussed in this Article, and conformal moments given by:
\begin{equation}
    \mathcal{C}_{n}(\xi, t) = \xi^{n} \int_{-1}^{1} \dx~C_{n}^{\nicefrac{3}{2}} \left( \frac{x}{\xi} \right) H(x, \xi, t) \,,
\end{equation}
where $C_{n}^{\nicefrac{3}{2}} (x)$ are Gegenbauer polynomials:
\begin{equation}
    C_{n}^{\nicefrac{3}{2}}(x) = 
    \sum_{\substack{k=0 \\ \mathrm{even} }}^{n} c_{n,k} x^{n-k} \,,
\end{equation}
with coefficients 
\begin{equation}
    c_{n,k} = (-1)^{\nicefrac{k}{2}}\frac{\Gamma(n-\nicefrac{k}{2}+3/2)}{\Gamma(3/2)(k/2)!(n-k)!}2^{n-k} \,.
\end{equation}
A finite number of Mellin moments is needed to express a single conformal moment of a fixed order, 
\begin{equation}
    \mathcal{C}_{n}(\xi, t) =\sum_{\substack{k=0 \\ \mathrm{even}}}^{n}
    c_{n,k}\xi^{k}
    \mathcal{A}_{n-k}(\xi, t) \,,
\end{equation} and \emph{vice versa}, a finite number of conformal moments is needed to express a single Mellin moment,
\begin{gather}
    c_{0,0} \mathcal{A}_{0}(\xi, t) = \mathcal{C}_{0}(\xi, t) \,, \\
    c_{n,0} 
    \mathcal{A}_{n}(\xi, t) =
    \mathcal{C}_{n}(\xi, t) -
    \sum_{\substack{k=2 \\ \mathrm{even}}}^{n}
    c_{n,k}\xi^{k}
    \mathcal{A}_{n-k}(\xi, t) \,.
\end{gather}
The conformal moments conveniently diagonalise the evolution equations at  leading-order (LO) \cite{Diehl:2003ny}, but they also appear in the modelling of GPDs, like for instance in Ref. \cite{Kumericki:2009uq}. 

GPDs may be equivalently \cite{Muller:2014wxa} defined in the double distribution representation of $\beta$ and $\alpha$ variables, which is related to $(x, \xi)$-space by the Radon transform:
\begin{align}
    f(x,\xi) = \int \mathrm{d}\Omega\,  F(\beta, \alpha) \,,
    \label{eq:theory:dd}
\end{align} 
where $\mathrm{d}\Omega =\mathrm{d}\beta\,\mathrm{d}\alpha\,\delta(x-\beta-\alpha\xi)$ and $|\alpha|+|\beta|\leq 1$. Working in $(\beta, \alpha)$-space allows us to relatively easily fulfil both the reduction to PDFs and the polynomiality property. However, any projection of experimental observables using GPD models defined in that space, or \emph{vice versa}, any attempt to find free parameters of such models from experimental data, requires either the Radon transform or its inverse counterpart. This is typically done numerically, which severely slows down computations. In some cases the Radon transform can be performed analytically, but it requires a rather simple form of double distributions and may be numerically unstable due to delicate cancellations of small numbers, typically occurring at small $\xi$ (see \eg the $\mathcal{O}\left(\xi^{-5}\right)$ factor in Eq.~(27) of Ref.~\cite{Goloskokov:2006hr}). The inverse Radon transform poses a challenge of its own, as reported for instance in Ref.~\cite{Chouika:2017dhe}. 

We close this section with the remark that the core of this work is related to preserving proper forward limits, polynomiality and positivity of GPDs models. This primarily does not involve the $t$ variable, which in general is not relevant in our study and as in other analyses of this type will suppressed for brevity. In summary, we are left with the problem of modeling GPDs in 2D spaces: either $(x, \xi)$ or $(\beta, \alpha)$.

%% file: sec_xxi.tex
\section{Modelling in $(x, \xi)$-space}
\label{sec:xxispace}

\subsection{Principles of modelling}
\label{sec:xxispace:principles}

The polynomiality property suggests that the $A_{n, j}$ coefficients introduced in Eq.~\eqref{eq:theory:polynomiality} provide a convenient basis to describe GPDs in a flexible modelling based on multiple parameters. The coefficients are not expected to be correlated, making them true degrees of freedom for GPDs. Another reason to choose this basis is the relation between Mellin and conformal moments indicated in Sect.~\ref{sec:theory}. This relation allows one to perform the LO evolution of GPDs in a fast and straightforward way. Finally, the strategy advocated here is a convenient way to incorporate calculations of Mellin moments performed by lattice QCD in the modelling of GPDs.

Because there are infinitely many $A_{n, j}$ coefficients, for a practical use we only consider a subset of them. This is justified by the fact that the evaluation of higher order Mellin moments becomes increasingly sensitive to the numerical noise: if $H(x, \xi)$ is a bounded summable function, its Mellin moment $\mathcal{A}_n(\xi)$ tends to zero when $n$ gets to infinity, in which case it starts to drown in the noise in any evaluation. This prevents the extraction of high order Mellin moments from experimental data without adding prior assumptions.  Evaluation on the lattice also suffers from increased uncertainty at higher orders. Therefore, even if in principle the full series of Mellin moments is required to unambiguously reconstruct a function, this situation almost never appears in practical situations. The general case to consider is actually when only a finite number of Mellin moments are known. An extra regularization is thus necessary to restore the uniqueness of the function defined by its Mellin moments (in the present case, this extra regularization is introduced by the choice of a polynomial basis or an ANN basis as described below). By linearity, we point out that this argument still holds true when dealing with conformal moments instead of Mellin moments.

In our modelling, we only let free the coefficients $A_{n,j}$ appearing in Mellin moments up to and including the order $n \leq N$. The value of $N$ is arbitrary and it controls the flexibility of models, just like the order of polynomials or the size of artificial neural networks used in modern parameterisations of PDFs. In other words, $N$ fixes the number of degrees of freedom. For even values of $N$, which for brevity we only consider in this study, this number is $(1+\nicefrac{N}{2})(2+\nicefrac{N}{2})-1$.

\begin{figure}[!ht]
\begin{center}
\includegraphics[width=0.45\textwidth]{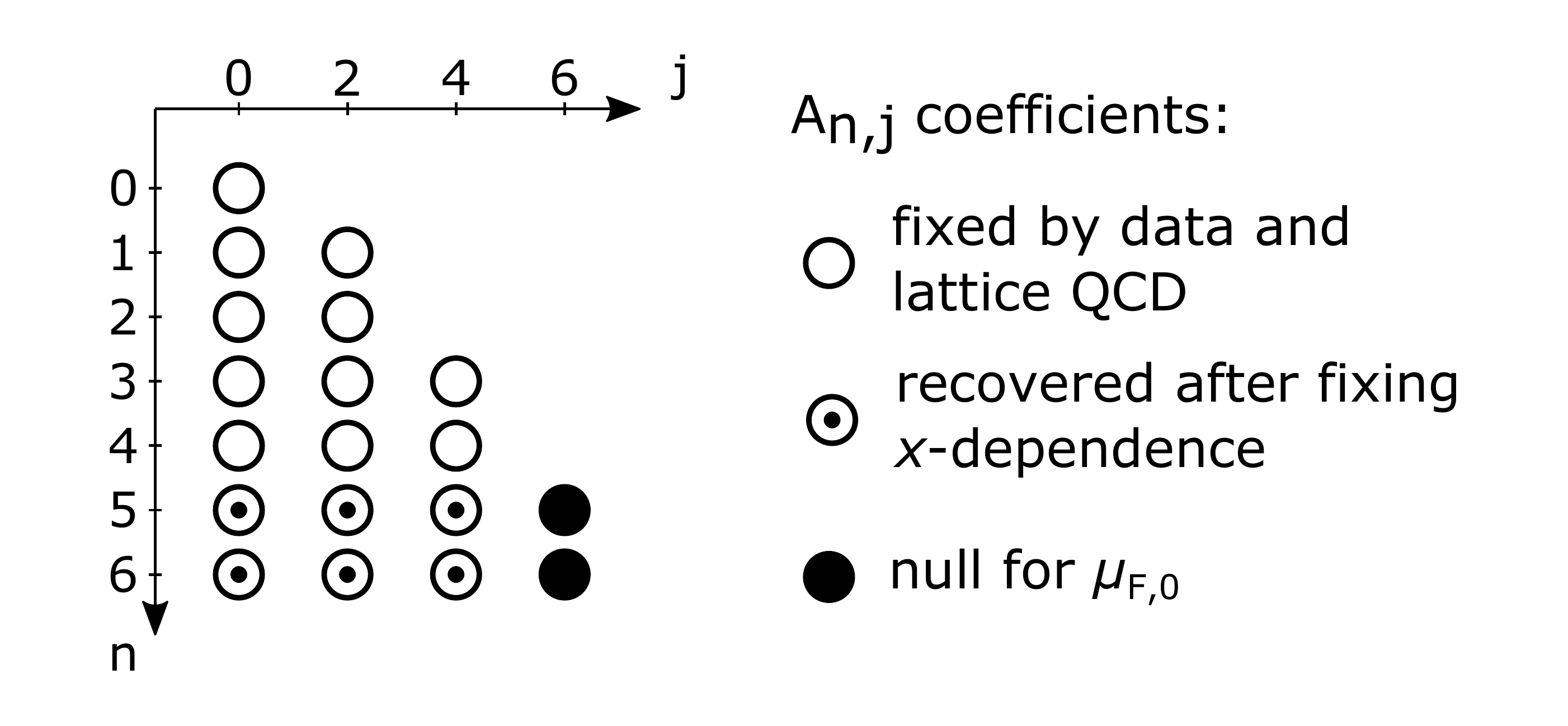}
\caption{The sketch indicates three groups of the polynomiality expansion coefficients, $A_{n,j}$, appearing in the modelling after fixing the truncation parameter to $N=4$. The coefficients that we assume to be constrained by data and lattice QCD are denoted by $\fullmoon$, those that are null, but only at the initial factorisation scale are denoted by $\newmoon$, while those calculable from $\fullmoon$ after fixing the recovery of $x$-dependence are denoted by $\odot$.}
\label{fig:xxispace:N_sketch}
\end{center}
\end{figure}

The coefficients $A_{n,j}$ for $n > N$ and $j \leq N$ are recovered from those for $n \leq N$ by a specific reconstruction procedure of the $x$-dependence that is discussed in the following. On the contrary, we assume that the coefficients $A_{n,j}$ vanish for $n > N$ and $j > N$. Note that this is true at $\mu_{F,0}$ where we are defining the model. For $\mu_F \neq \mu_{F,0}$, evolution will make these once null coefficients to have non-zero values, but only for $j \leq n+\textrm{mod}(n, 2)$ since evolution preserves the polynomiality property. We summarise this paragraph with a sketch shown in Fig. \ref{fig:xxispace:N_sketch}. 

The reconstruction of the $x$-dependence from a number of $A_{n, j}$ coefficients is the \textit{moment problem} known in mathematics. Let us address it by expressing GPDs in a way suggested by the polynomiality property:
\begin{equation}
H(x,\xi) = \sum_{\substack{j=0 \\ \mathrm{even}}}^{N}f_{j}(x)\xi^{j} \,.
\label{eq:xxispace:master}
\end{equation}
Here, $f_{j}(x)$ is a function of $x$ that satisfies the following relation to Mellin coefficients: 
\begin{equation}
 \int_{-1}^{1}\dx x^{n} f_{j}(x) = A_{n,j} \,,
\label{eq:xxispace:Anj_from_fjx}    
\end{equation}
where $A_{n,j} = 0$ for $j > n+\mod(n, 2)$.
Since $H(x = \pm 1, \xi) = 0$ and considering the polynomial form of Eq.~\eqref{eq:xxispace:master}, $f_{j}(x)$ has to vanish at the end points:
\begin{equation}
f_{j}(-1) = f_{j}(1) = 0 \,.
\label{eq:xxispace:border_fjx}
\end{equation}

Equations \eqref{eq:xxispace:Anj_from_fjx} and \eqref{eq:xxispace:border_fjx} are essential to constrain $f_{j}(x)$ and therefore to obtain valid GPD models. In the following we will try two bases for $f_{j}(x)$, \ie two forms of this function: one based on monomials, which has already been studied in the literature (see Ref.~\cite{Muller:2014wxa} and  references therein), and one inspired by the ANN technique. A given basis fixes the bridge between the picture of GPDs given by Mellin moments ($\xi$-dependent integrals over $x$) and the picture where GPDs are explicitly given as functions of $x$ and $\xi$. We note that the selection of a given basis and the necessity of keeping a finite $N$ introduces some model-dependency. We however expect this bias to become arbitrarily small for sufficiently large $N$.

It is important to note that GPDs may exhibit a singularity at $x=\xi=0$, which corresponds to the singularity of PDF at $x=0$. However, this singularity does not affect the evaluation of Mellin moments. Since here we are modeling GPDs with a polynomial in $\xi$, see Eq.~\eqref{eq:xxispace:master}, we are not able to effectively build a model that exhibits the singularity at only $x=\xi=0$ (and not at $x=0$, $\xi \neq 0$). It is a general flaw of this type of modelling, which motivated us to find ANN-inspired GPDs in $(\beta, \alpha)$-space, which we present in Sect.~\ref{sec:baspace}. Still, a direct modelling of $(x,\xi)$-space can be used for a subset of non-singular GPD models, like the pion model \cite{Chouika:2017rzs,Chavez:2021llq} used for a demonstration in Sect.~\ref{sec:xdep:demo}. In addition, one can consider a reference GPD model, $H_{0}(x,\xi)$, with a correct $H_{0}(x,\xi)/q(x)$ behaviour for $\xi \rightarrow 0$, and constrain a polynomial model, $H_{1}(x,\xi)$, such that $H_{0}(x,\xi) + H_{1}(x,\xi)$ correctly reproduces $\xi > 0$ data and \eg lattice results for Mellin moments. This kind of hybrid modelling may be plausible for phenomenology, as models defined in $(x,\xi)$-space can be directly compared to data, and therefore their usage can severely speed up minimization procedures. 

\subsection{Polynomial basis}
\label{sec:xxispace:polynomail}

According to the Stone-Weierstrass theorem, any continuous function defined on a compact set can be uniformly approximated by a polynomial to any desired degree of precision. Applying this theorem to the two-dimensional function $H(x, \xi)$ suggests to express $f_{j}(x)$ that is defined in the $|x| \leq 1$ interval by 
\begin{equation}
f_{j}(x) = \sum_{i=0}^{N+2} w_{i,j} x^{i} \,.
\label{eq:xxispace:monomial_basis}
\end{equation}
Here, $w_{i,j}$ are coefficients multiplying the $x^{i}\xi^{j}$ monomials in the global expression of $H(x, \xi)$. The order of polynomial \eqref{eq:xxispace:monomial_basis} is $N+2$, and it is the minimal order required to satisfy both Eqs. \eqref{eq:xxispace:Anj_from_fjx} and \eqref{eq:xxispace:border_fjx}. We note that the order may be higher, which provides extra degrees of freedom available in the modelling. 

Mellin coefficients can be evaluated from monomial ones in the following way:
\begin{align}
\displaystyle\mathop{\forall}_{\mathrm{even}~n}
A_{n, j} = &
\sum_{\substack{i=0 \\ \mathrm{even}}}^{N+2} w_{i, j} \int_{-1}^{1} \dx x^{i+n} = \sum_{\substack{i=0 \\ \mathrm{even}}}^{N+2} w_{i, j} \frac{2}{1+i+n}  \,,
\label{eq:xxispace:Anj_from_monomial_even} \\
\displaystyle\mathop{\forall}_{\mathrm{odd}~n}
A_{n, j} = &
\sum_{\substack{i=1 \\ \mathrm{odd}}}^{N+1} w_{i, j} \int_{-1}^{1} \dx x^{i+n} = \sum_{\substack{i=1 \\ \mathrm{odd}}}^{N+1} w_{i, j} \frac{2}{1+i+n} \,,
\label{eq:xxispace:Anj_from_monomial_odd}
\end{align}
where we remind that $N$ is even, and where Eq.~\eqref{eq:xxispace:border_fjx} yields:
\begin{equation}
\sum_{i = 0}^{N+2}w_{i,j} = \sum_{i = 0}^{N+2}(-1)^{i}w_{i,j} = 0 \,.
\label{eq:xxispace:border_monomial}
\end{equation}

To find the opposite relations we need to solve Eqs.~\eqref{eq:xxispace:Anj_from_monomial_even}, \eqref{eq:xxispace:Anj_from_monomial_odd} and \eqref{eq:xxispace:border_monomial} for $w_{i, j}$. We express all these equations in the matrix form: 
\begin{equation}
\vec{A}_{j}=\mathcal{C}\cdot\vec{w}_{j}\,.
\label{eq:xxispace:matrix_equation}
\end{equation}
Here, $\vec{A}_{j}=\left(A_{0,j}, \ldots, A_{n,j}, \ldots, A_{N,j} , 0, 0\right)^{\mathrm{T}}$ and $\vec{w}_{j}=\left(w_{0,j}, \ldots, w_{i,j}, \ldots, w_{N+2,j} \right)^{\mathrm{T}}$, and the length of each vector is $N+3$. The matrix $\mathcal{C}$ is:
\begin{gather}
\mathcal{C} = 
\begin{pmatrix}
c_{0,0} & \ldots & c_{0,i} & \ldots & c_{0,N+2} \\
\ldots & \ldots &\ldots & \ldots & \ldots \\
c_{n,0} & \ldots & c_{n,i} & \ldots & c_{n,N+2} \\
\ldots & \ldots &\ldots & \ldots & \ldots \\
c_{N+2,0} & \ldots & c_{N+2,i} & \ldots & c_{N+2,N+2}
\end{pmatrix} \,,
\end{gather}
where:
\begin{equation}
 c_{n,i}=
    \begin{cases}
     \displaystyle\frac{2}{1 + i + n} & n \leq N~\mathrm{and} ~\mathrm{even}~n+i \,, \\
      0 & n \leq N~\mathrm{and} ~\mathrm{odd}~n+i \,, \\
      (-1)^{i \, n} & \mathrm{otherwise} \,.
    \end{cases}   
\end{equation}
This matrix is only characterized by the parameter $N$. Therefore, as soon as this parameter is fixed the matrix is ready for the inversion (for instance numerically), and this can be done only once in the modelling procedure. 

We note that the presented realisation of GPDs in the monomial basis of minimal order $N+2$ naturally arises in the context of the so-called dual parameterisation \cite{Polyakov:2002wz} after its initial formal representation has been recast as a convergent series. This parameterisation is physically motivated by an infinite series of $t$-channel exchanges. We also note that any attempt of modelling GPDs in the $(x, \xi)$-space based on classical orthogonal polynomials of $x$, like Gegenbauer polynomials, will lead to the same conclusions as presented in this section. This is because any monomial of $x$ can be expressed by a finite number of classical orthogonal polynomials.

\subsection{Artificial neural network basis}
\label{sec:xdep:ann}

\begin{figure}[!ht]
\begin{center}
\includegraphics[width=0.45\textwidth]{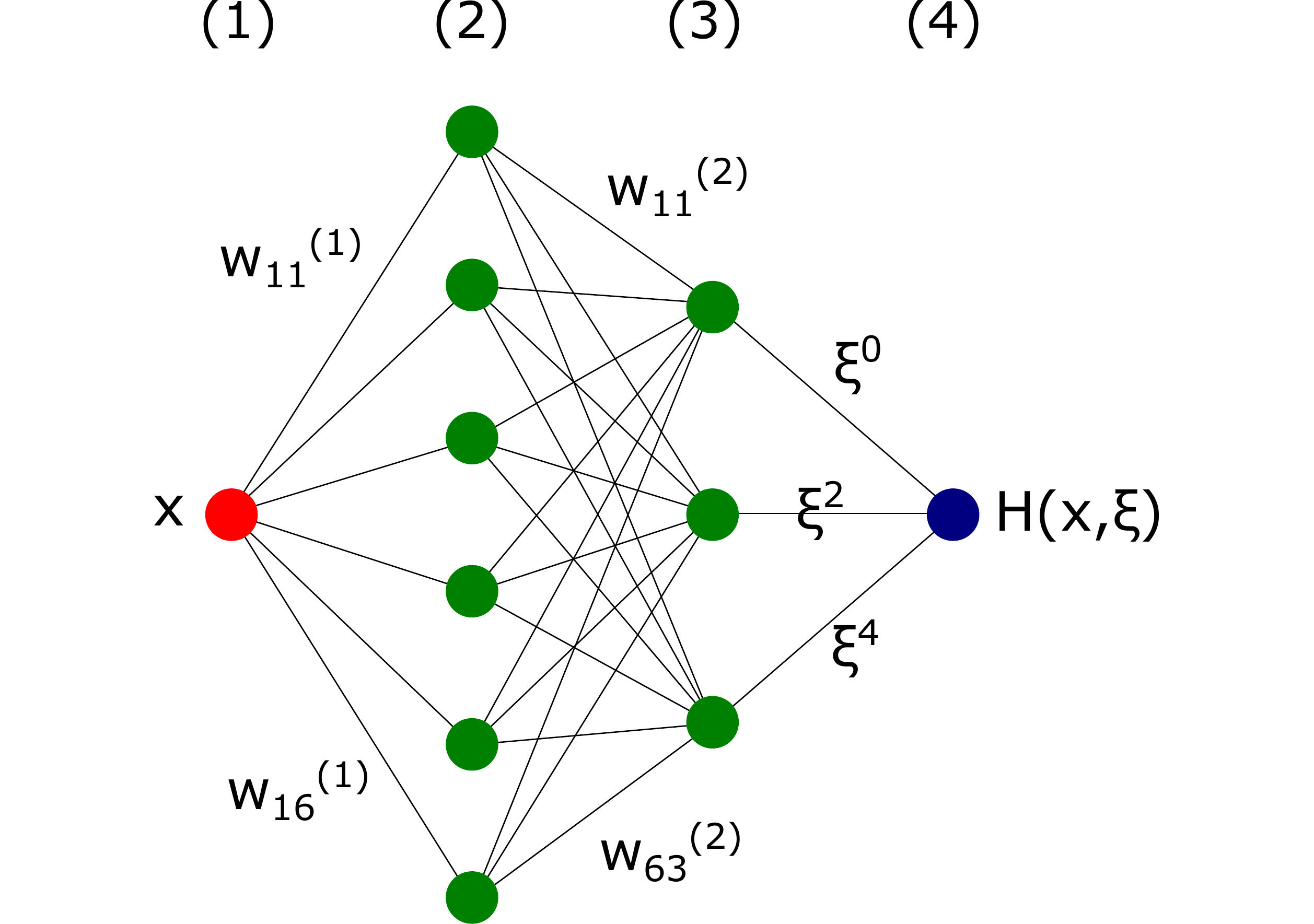}
\caption{Scheme of an exemplary artificial neural network used to represent a single GPD. The example is for the truncation parameter $N=4$.}
\label{fig:xdep:ann:scheme}
\end{center}
\end{figure}

In the case of ANNs, universal approximation theorems (see \eg Ref.~\cite{Cybenko}) for arbitrary width or depth of the neural network ensure that a large enough network is able to accurately represent any continuous function on a compact set. The graphical representation of an exemplary network used to construct valid GPD models is shown in Fig.~\ref{fig:xdep:ann:scheme}. 

The exemplary network is made out of four layers: input and output layers ($(1)$ and $(4)$, respectively), and two hidden layers ($(2)$ and $(3)$). The signal is processed through the network as follows:
\begin{enumerate}[label=\emph{\roman*})]
	\item The neuron in the input layer receives the value of $x$ and distributes it to all neurons in the first hidden layer (second layer of the network: $(2)$) via a set of connections. A single weight is associated to each of those connections that is denoted by $w_{1,i}^{(1)}$, where in our example $i = 1, \ldots, 6$. 
	\item Each neuron in the hidden and output layers processes the incoming  signal according to this equation:
	\begin{equation}
		o^{(l)}_{k} = \varphi^{(l)}_{k}\left(b_{k}^{(l)} + \sum_{i}o^{(l-1)}_{i}w^{(l-1)}_{i,k}\right) \,,
	\end{equation}
	where $o^{(l)}_{k}$ is the output of $k$-th neuron in $(l)$ layer, $\varphi^{(l)}_{k}(\cdot)$ is the activation function, $w^{(l-1)}_{i,k}$ is the weight associated to the connection between $i$-th neuron from $(l-1)$ layer and $k$-th neuron from $(l)$ layer, and $b_{k}^{(l)}$ is the bias parameter. We note that:
	\begin{gather}
		o_{1}^{(1)} \equiv x \,, \\
		o_{1}^{(4)} \equiv H(x,\xi) \,.
	\end{gather}
	\item To reproduce the form of Eq.~\eqref{eq:xxispace:master} we use the linear activation function for the neuron in the output layer:
	\begin{equation}
		\varphi^{(4)}_{1}(\cdot) = (\cdot) \,,
	\end{equation}
	where $(\cdot)$ denotes the function argument, and we introduce the $\xi$-dependence via weights associated to the connections linking the last hidden layer with the output layer:
	\begin{equation}
		w^{(3)}_{i,1} = \xi^{2(i-1)} \,.
	\end{equation}
	We also set the bias in the output neuron to:
	\begin{equation}
		b_{1}^{(4)} = 0 \,.
	\end{equation}
	This choice means that $f_{j}(x)$ associated to $\xi^j$, see Eq.~\eqref{eq:xxispace:master}, is effectively described by an ANN with a single hidden layer corresponding to the first hidden layer of the full network, and the output layer corresponding to one neuron in the second hidden layer of the full network.
\end{enumerate}

We are free to choose the activation function associated to the neurons in the hidden layers. For the purpose of further demonstration in the first hidden layer we use either the sigmoid function,
\begin{gather}
\varphi^{(2)}_{k}(\cdot) = \frac{1}{1+\exp\left(-(\cdot)\right)}\,, \label{eq:xxispace:act_sigmoid}
\end{gather} 
or the rectifier function (ReLU),
\begin{gather}
\varphi^{(2)}_{k}(\cdot) = (\cdot)\, \Theta(\cdot)\,, \label{eq:xxispace:act_relu}
\end{gather} 
where $\Theta(\cdot)$ is the Heaviside step function. In the second hidden layer the linear activation function will be used,
 \begin{gather}
\varphi^{(3)}_{k}(\cdot) = (\cdot)\,. \label{eq:xxispace:act_linear}
\end{gather} 
This choice allows us to analytically evaluate the Mellin coefficients using:
\begin{strip}
\newline
\rule{\textwidth}{\arrayrulewidth}
\begin{align}
\Phi^{(2)}_{k}(n)= 
\int_{-1}^{1}\dx x^{n}o^{(2)}_{k} =
\int_{-1}^{1}\dx x^{n}\varphi^{(2)}_{k}\left({b^{(2)}_{k}} + {w^{(1)}_{1,k}}x\right) 
\end{align}
for the sigmoid function Eq.~\eqref{eq:xxispace:act_sigmoid}:
\begin{align}
\phantom{\Phi^{(2)}_{k}(n)}=
(-1)^{n+1}\sum_{i=0}^{n} 
\left({w^{(1)}_{1,k}}\right)^{-i-1} \frac{\Gamma(1+n)}{\Gamma(1+n-i)} \left((-1)^{n+i} \mathrm{Li}_{i+1}\left(-e^{{b^{(2)}_{k}}+{w^{(1)}_{1,k}}}\right) -
\mathrm{Li}_{i+1}\left(-e^{{b^{(2)}_{k}}-{w^{(1)}_{1,k}}}\right)\right) \,, 
\end{align}
or, for the ReLU function Eq.~\eqref{eq:xxispace:act_relu}:
\begin{gather}
\phantom{\Phi^{(2)}_{k}(n)}=
\begin{cases}
\displaystyle\frac{\left((-1)^n+1\right)}{n+1}b^{(2)}_{k}-\frac{\left((-1)^n-1\right)}{n+2}w^{(1)}_{1,k} & \mathrm{for}~\left|b^{(2)}_{k}/w^{(1)}_{1,k}\right| \geq 1 ~\mathrm{and}~ s_{b/w} \neq s_{w}\,, \\
0 & \mathrm{for}~\left|b^{(2)}_{k}/w^{(1)}_{1,k}\right| \geq 1 ~\mathrm{and}~ s_{b/w} = s_{w}\,, \\
\displaystyle
\frac{b^{(2)}_{k}w^{(1)}_{1,k}\left(-\displaystyle\frac{b^{(2)}_{k}}{w^{(1)}_{1,k}}\right)^{n+1}+s_{w}^{n+1} w^{(1)}_{1,k} \left( w^{(1)}_{1,k}(n+1)+s_{w}b^{(2)}_{k}(n+2)\right)}{(n+1) (n+2)w^{(1)}_{1,k}} & \mathrm{otherwise}\,,
\end{cases}
\end{gather}
where:
\begin{gather}
s_{b/w} = \mathrm{sgn}\left({b^{(2)}_{k}/w^{(1)}_{1,k}}\right),~
s_{w} = \mathrm{sgn}\left(w^{(1)}_{1,k}\right)\,,
\end{gather}
independently of the choice of the activation function in the first hidden layer:
\begin{align}
\Phi^{(3)}_{k}(n) & =
\int_{-1}^{1}\dx x^{n}o^{(3)}_{k} = 
\int_{-1}^{1}\dx x^{n}\varphi^{(3)}_{k}\left(b^{(3)}_{k} + \sum_{i}w^{(2)}_{i,k}\varphi^{(2)}_{i}\left(w_{1,i}^{(1)}x+b_{i}^{(2)}\right)\right)  \nonumber \\
& = \mod(n+1,2)\frac{2}{n+1}b^{(3)}_{k}
+ \sum_{i}w^{(2)}_{i,k}\Phi^{(2)}_{i}(n) \,,
\end{align}
\rule{\textwidth}{\arrayrulewidth}
\end{strip}
where $\mathrm{Li}_i(x)$ is the polylogarithm and
\begin{equation}
\Phi^{(3)}_{1+j/2}(n) \equiv A_{n,j}\,.
\end{equation}
These coefficients depend linearly on $b_{k}^{(3)}$ and $w_{i,k}^{(2)}$, so the later can be found from a number of Mellin moments by solving the system of equations similar to that of Eq.~\eqref{eq:xxispace:matrix_equation}. Here, $\vec{A}_{j}=\left(A_{0,j}, \ldots, A_{i,j}, \ldots, A_{N,j} , 0, 0\right)^{\mathrm{T}}$ is the same as for the polynomial basis, while $\vec{w}_{j}=\left(b_{1+j/2}^{(3)},  w_{1,1+j/2}^{(2)}, \ldots, w_{N+2,1+j/2}^{(2)} \right)^{\mathrm{T}}$ and
\begin{equation}
 c_{n,i}=
    \begin{cases}
        2\, \mod(n+1, 2) / (n+1) & n \leq N ~\mathrm{and}~ i = 0 \,, \\
        \Phi_{i}^{(2)}(n) & n \leq N ~\mathrm{and}~ i > 0 \,, \\
        1 & n > N ~\mathrm{and}~ i = 0 \,, \\
      \varphi_{i}^{(2)}\left(b_{i}^{(2)} + (-1)^n w_{1,i}^{(1)}\right) & \mathrm{otherwise}  \,.
    \end{cases}   
\end{equation}
The size of the last hidden layer (third layer of the network: $(3)$) is $N/2+1$, while to ensure both the polynomiality and vanishing at $|x|=1$ the size of the fist hidden layer (second layer of the network: $(2)$) must be at least $N+2$.

In our approach the coefficients of the first hidden layer, $w_{1,i}^{(1)}$ and $b_{i}^{(2)}$, are not constrained by the Mellin moments. Values of these coefficients can be taken random, or they can be fixed with further criteria, \eg to obtain the best possible agreement between the input PDF and the resulting GPD in $x$-space. The sensitivity on the choice of $w_{1,i}^{(1)}$ and $b_{i}^{(2)}$ coefficients becomes small for large N.

\subsection{Demonstration of models}
\label{sec:xdep:demo}

\begin{figure*}[!ht]
\begin{center}
\includegraphics[width=0.35\textwidth]{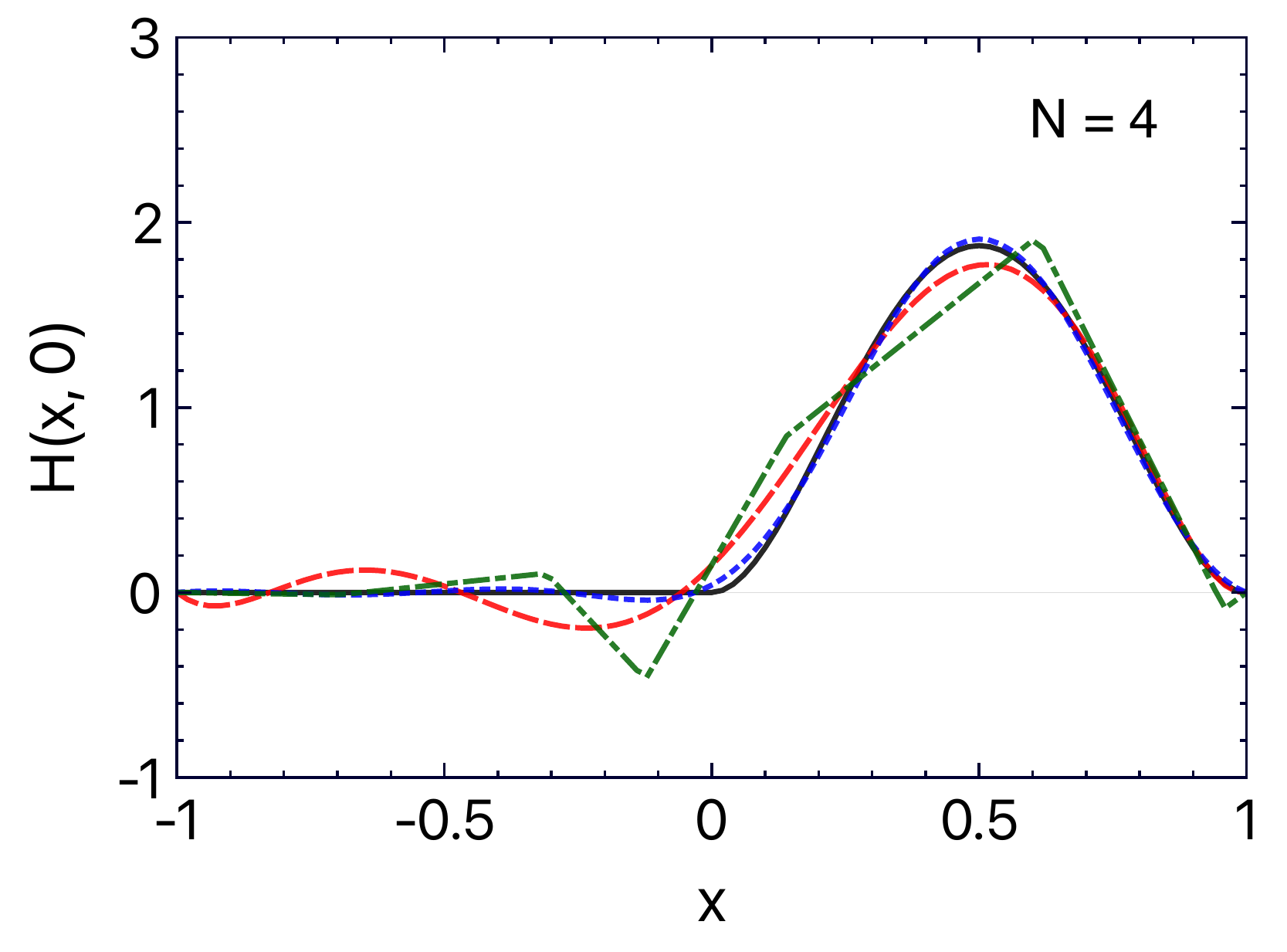}
\includegraphics[width=0.35\textwidth]{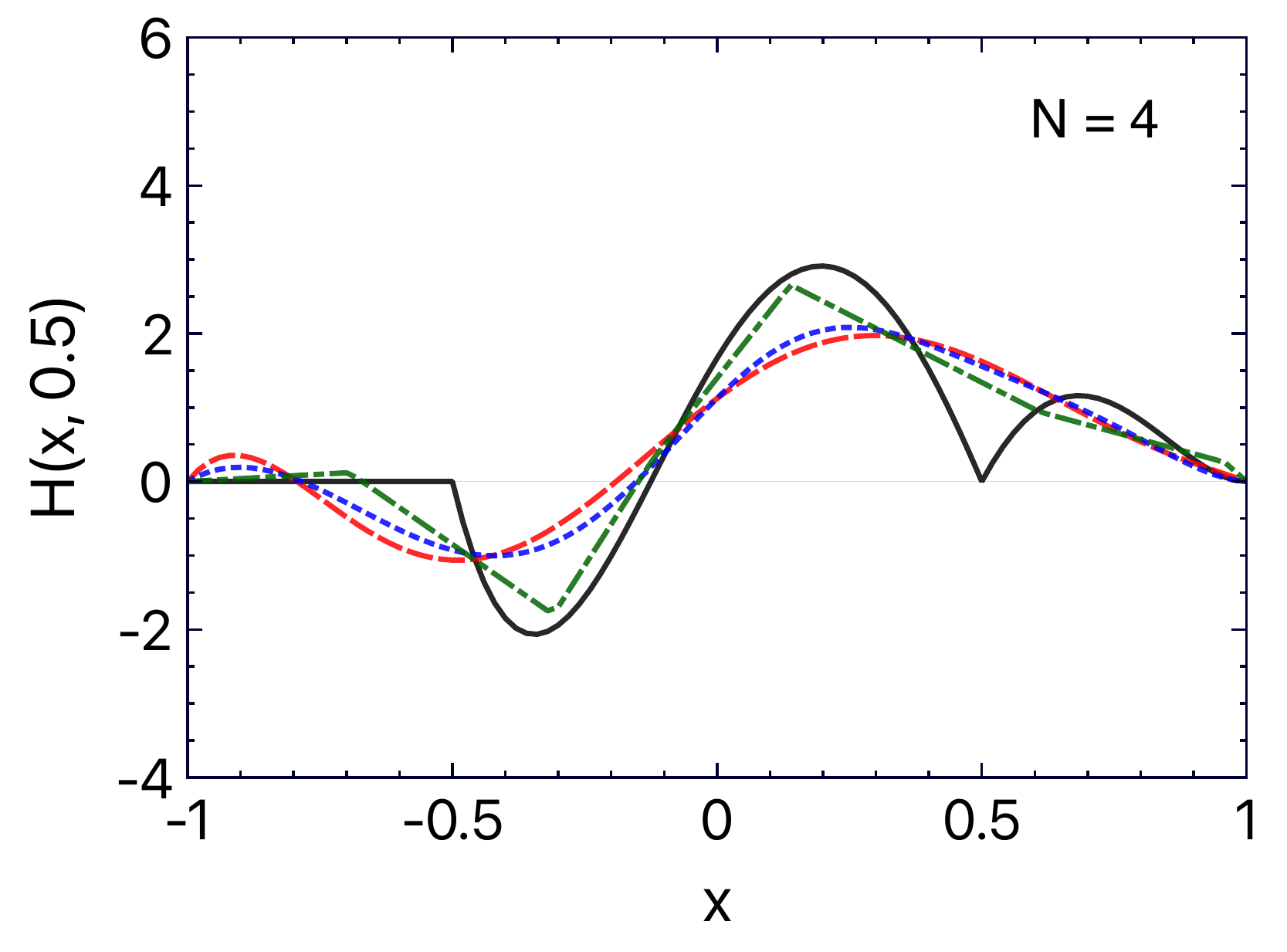}
\includegraphics[width=0.35\textwidth]{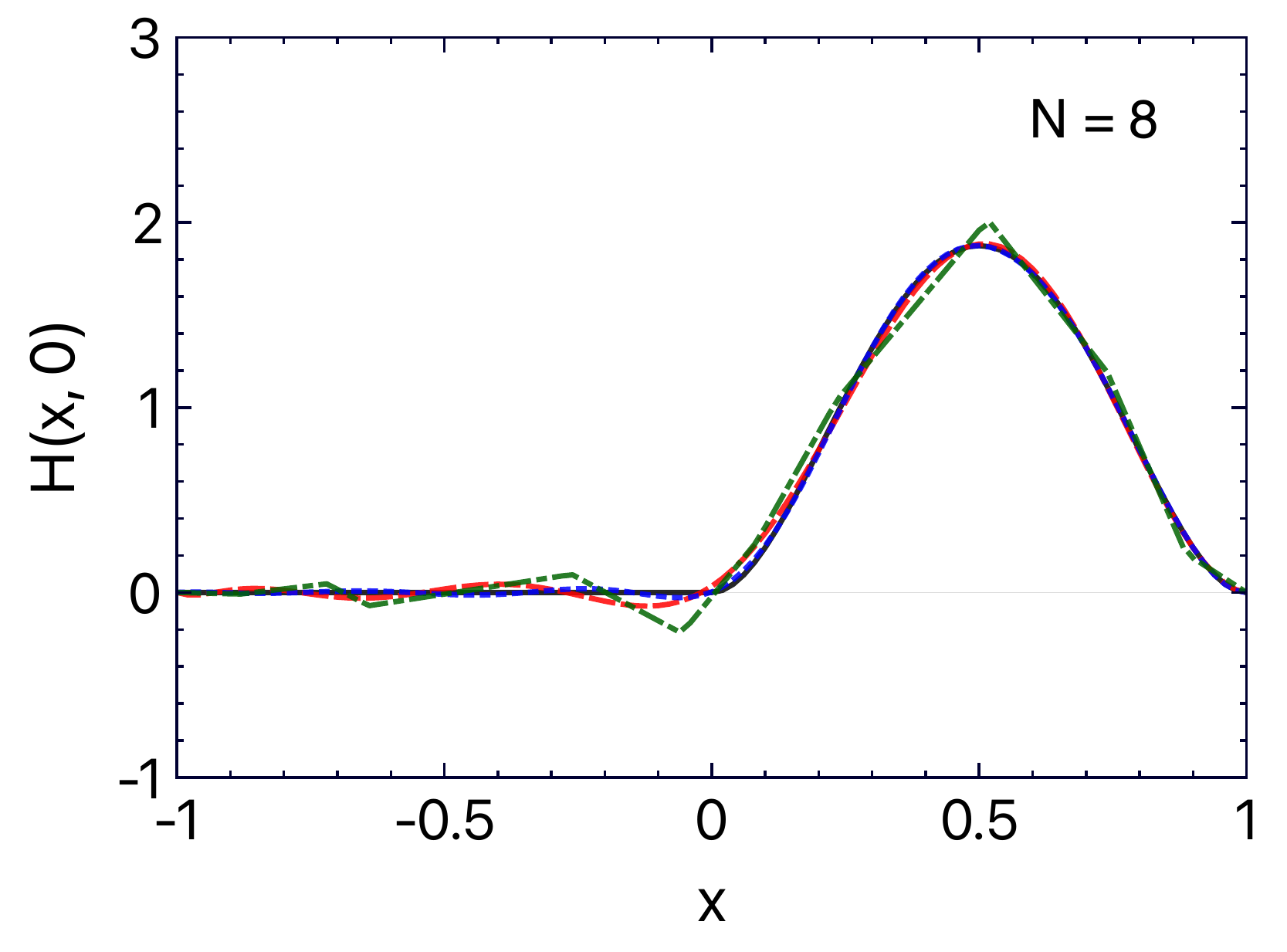}
\includegraphics[width=0.35\textwidth]{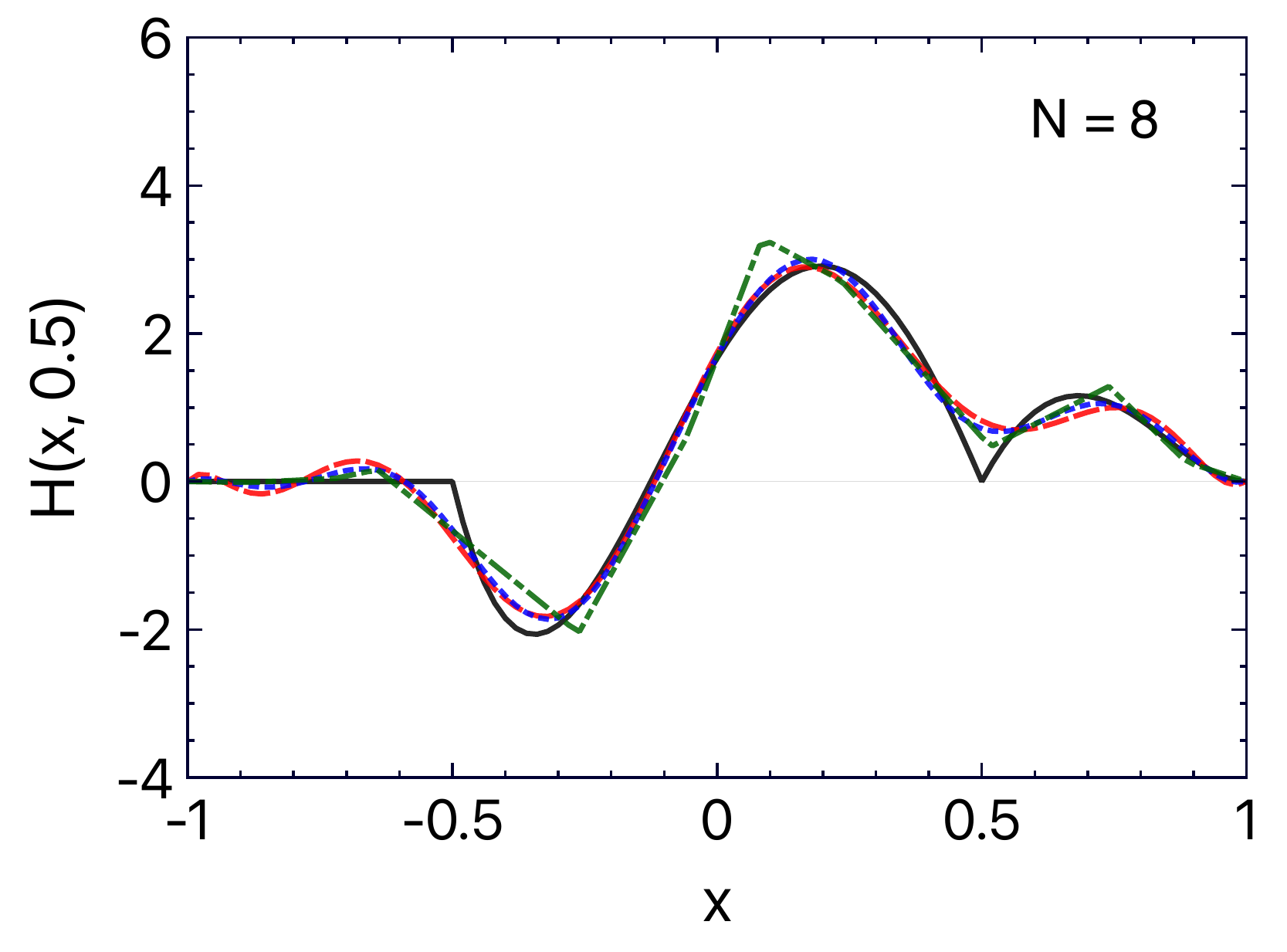}
\caption{The pion model \cite{Chouika:2017rzs,Chavez:2021llq} (black solid line) given by Eq.~\eqref{eq:xdep:demo:pionModel} reconstructed from either five ($N=4$, top row) or nine ($N=8$, bottom row) of its first Mellin moments as a function of $x$ for either $\xi=0$ (left column) or $\xi=0.5$ (right column). The reconstruction is done using a polynomial basis (red dashed line) or an ANN basis with either sigmoid activation function (blue dotted line) or ReLU activation function (green dash-dotted line).}
\label{fig:xdep:demo}
\end{center}
\end{figure*}

We illustrate the modelling presented in this section with the help of the following pion GPD \cite{Chouika:2017rzs,Chavez:2021llq}:
\begin{align}
    H_{\pi}(x,\xi) & = \Theta(x - |\xi|) \, \frac{30 (1 - x)^2 (x^2 - \xi^2)}{(1 - \xi^2)^2} \nonumber \\
   &  + \Theta(|\xi| - |x|) \, \frac{15 (1 - x) (\xi^2 - x^2) (x + 2x\xi + \xi^2)}{2\xi^3 (1 + \xi)^2} \,.
\label{eq:xdep:demo:pionModel}
\end{align}
We fix the truncation parameter $N$ to either $4$ or $8$, and we evaluate the corresponding $A_{n, j}$ coefficients from $H_{\pi}(x,\xi)$. We use those coefficients to reconstruct $H_{\pi}(x,\xi)$ using either the polynomial or ANN basis (with either the sigmoid or ReLU activation functions). This way we check how well one may reconstruct the underlying GPD only knowing a number of its Mellin moments. The result is shown in Fig.~\ref{fig:xdep:demo} for $\xi=0$ and $\xi=0.5$.

In this Article we only show the most elementary use of the modelling based on Eq.~\eqref{eq:xxispace:master}. However several possible extensions or modifications are possible. In particular, one may extend the sum in Eq.~\eqref{eq:xxispace:monomial_basis} (for polynomial basis) or the number of neurons in the first hidden layer (for ANN basis) to introduce more free parameters that can be ``spent'' for a better description of the $x$-dependence and of higher Mellin coefficients for a given power of $\xi$. One can also try the modelling with an explicit PDF contribution:
\begin{equation}
    H(x,\xi) = q(x) + \sum_{\substack{j=2 \\ \mathrm{even}}}^{N}f_{j}(x)\xi^{j} \,,
\end{equation}
which automatically fixes all $A_{n,0}$ coefficients. If needed, vanishing at $|x|=|\xi|$ can be enforced in this way: 
\begin{equation}
    H(x,\xi) = (x^2-\xi^2)\sum_{\substack{j=0 \\ \mathrm{even}}}^{N}f_{j}(x)\xi^{j} \,,
\end{equation}
which typically causes an oscillatory behaviour along the $x$-axis. One may also introduce a separate D-term \cite{Polyakov:1999gs}:
\begin{equation}
    H(x,\xi) = \textrm{sgn}(\xi) D(x/\xi) + \sum_{\substack{j=0 \\ \mathrm{even}}}^{N}f_{j}(x)\xi^{j}\,,
\end{equation}
here denoted by $D(x/\xi)$, which will contribute to $A_{n,n+1}$ coefficients. This separate contribution allows to reproduce GPDs, which are not smooth functions and usually exhibit a kink at $|x|=|\xi|$.

In our simple demonstration we do not check if the models fulfil the positivity constraint. With the pion model given by Eq.~\eqref{eq:xdep:demo:pionModel} used as a benchmark it is particularly difficult. From Eq.~\eqref{eq:theory:positivity} we see that if the PDF vanishes for $x < 0$, the corresponding GPD must vanish as well for $x < -|\xi|$. We are not able to achieve this behaviour easily via our modelling of Mellin moments. However, we note that with a different benchmark model it should be possible to use a numerical enforcement of positivity. This method will be introduced in the next section of this Article.

%% file: sec_ba.tex
\section{Modelling in $(\beta, \alpha)$-space}
\label{sec:baspace}

\subsection{Principles of modelling}
\label{sec:baspace:principles}

The advantage of modelling GPDs using double distributions is a natural provision of polynomiality by the Radon transform. In addition, a proper reduction to PDFs can be achieved easily, with a welcome possibility of keeping the singularity at $x=0$ for only $\xi=0$. The GPD support $x \in [-1, +1]$ is also naturally encoded in the double distribution support $|\alpha|+|\beta| \leq 1$.

We are modelling the GPD $H(x,\xi)$ with three contributions, each one being the subject to the Radon transform shown in Eq.~\eqref{eq:theory:dd}: 
\begin{equation}
    (1-x^2)  F_{C}(\beta, \alpha) + 
    (x^2 - \xi^2) F_{S}(\beta, \alpha) + 
    \xi F_{D}(\beta, \alpha)
    \label{eq:baspace:principles:mainAnsatz}
\end{equation}
The first term, $(1-x^2)F_{C}(\beta, \alpha)$ gives the ``classical'' contribution reproducing both the forward limit and the cross-over line $\xi = x$ \cite{Musatov:1999xp}. This term is crucial for the comparison of GPD models with experimental data available today. The second term, $(x^2 - \xi^2)F_{S}(\beta, \alpha)$, gives a ``shadow'' contribution, which vanishes for both $\xi = 0$ and $\xi = x$ \cite{Bertone:2021yyz}. As we will see, its inclusion is important for a proper estimation of model uncertainties when GPDs are constrained by a sparse set of data. The third term, $\xi F_{D}(\beta, \alpha)$, only contributes to the D-term, which is accessible in analyses of amplitudes for exclusive processes (see \eg Ref.~\cite{Dutrieux:2021nlz}). The inclusion of this term gives an extra flexibility to the model required to reproduce all $x$-moments of GPDs. 

We note that for singlet GPDs, for which even Mellin moments vanish, the prefactors $x^2$ and $\xi^2$ in Eq.~\eqref{eq:baspace:principles:mainAnsatz} do not break the polynomiality property. Indeed, for odd Mellin moments of order $n$, double distributions being the subject to the Radon transform shown in Eq.~\eqref{eq:theory:dd} give contributions up to the $A_{n,n-1}$ coefficient only -- since the $A_{n,n}$ coefficient is zero for parity reasons. Therefore, adding either an $x^2$ or $\xi^2$ factor only brings contributions up to $A_{n,n+1}$, which is compatible with polynomiality. 

\subsection{$F_{C}(\beta, \alpha)$ contribution}
\label{sec:baspace:Fu}

We factorise $F_{C}(\beta, \alpha)$ into the forward limit, $f(\beta)$, which is well-known, at least for the GPD $H(x,\xi)$, and the profile function $h_{C}(\beta, \alpha)$:
\begin{equation}
F_{C}(\beta, \alpha) = f(\beta) h_{C}(\beta, \alpha)\frac{1}{1-\beta^{2}} \,.
\label{eq:baspace:ddfactorisationFu}
\end{equation}
The prefactor $(1-x^2)/(1-\beta^2)$ arising from Eqs.~(\ref{eq:baspace:principles:mainAnsatz}) and (\ref{eq:baspace:ddfactorisationFu}) turns out to be convenient to preserve positivity with an ANN-based modelling of $h_{C}(\beta, \alpha)$. Since in this study we are only interested in the singlet contribution we fix:
\begin{equation}
f(\beta) = \mathrm{sgn}(\beta) q(|\beta|) \,.
\label{eq:baspace:pdf}
\end{equation}
The profile function must fulfil the following set of properties to ensure the correct behaviour of the GPD in $(x, \xi)$-space: 
\begin{enumerate}[label=\emph{\roman*})]
\item even parity w.r.t. the $\beta$ variable, 
\begin{equation}
h_{C}(\beta, \alpha) = h_{C}(-\beta, \alpha) \,,
\end{equation}
to keep the whole GPD an odd function of $x$. 
\item even parity w.r.t. the $\alpha$ variable,
\begin{equation}
h_{C}(\beta, \alpha) = h_{C}(\beta, -\alpha) \,,
\end{equation}
to keep the whole GPD an even function of $\xi$, and therefore to hold time reversal symmetry.
\item the following normalisation: 
\begin{equation}
\int_{-1+\beta}^{1-\beta} \mathrm{d}\alpha h_{C}(\beta, \alpha) = 1 \,,
\label{eq:ba:normTo1}
\end{equation}
to ensure the proper reduction to the PDF at $\xi = 0$. 
\item vanishing at the edge of the support region,
\begin{equation}
h_{C}(\beta, \alpha) = 0~~~\mathrm{for}~~~|\beta|+|\alpha| = 1 \,,
\end{equation}
to avoid any singular behaviour, except at the $x=\xi=0$ point, and to help enforcing the positivity property at $x \approx 1$.
\end{enumerate}

\begin{figure}[!ht]
\begin{center}
\includegraphics[width=0.45\textwidth]{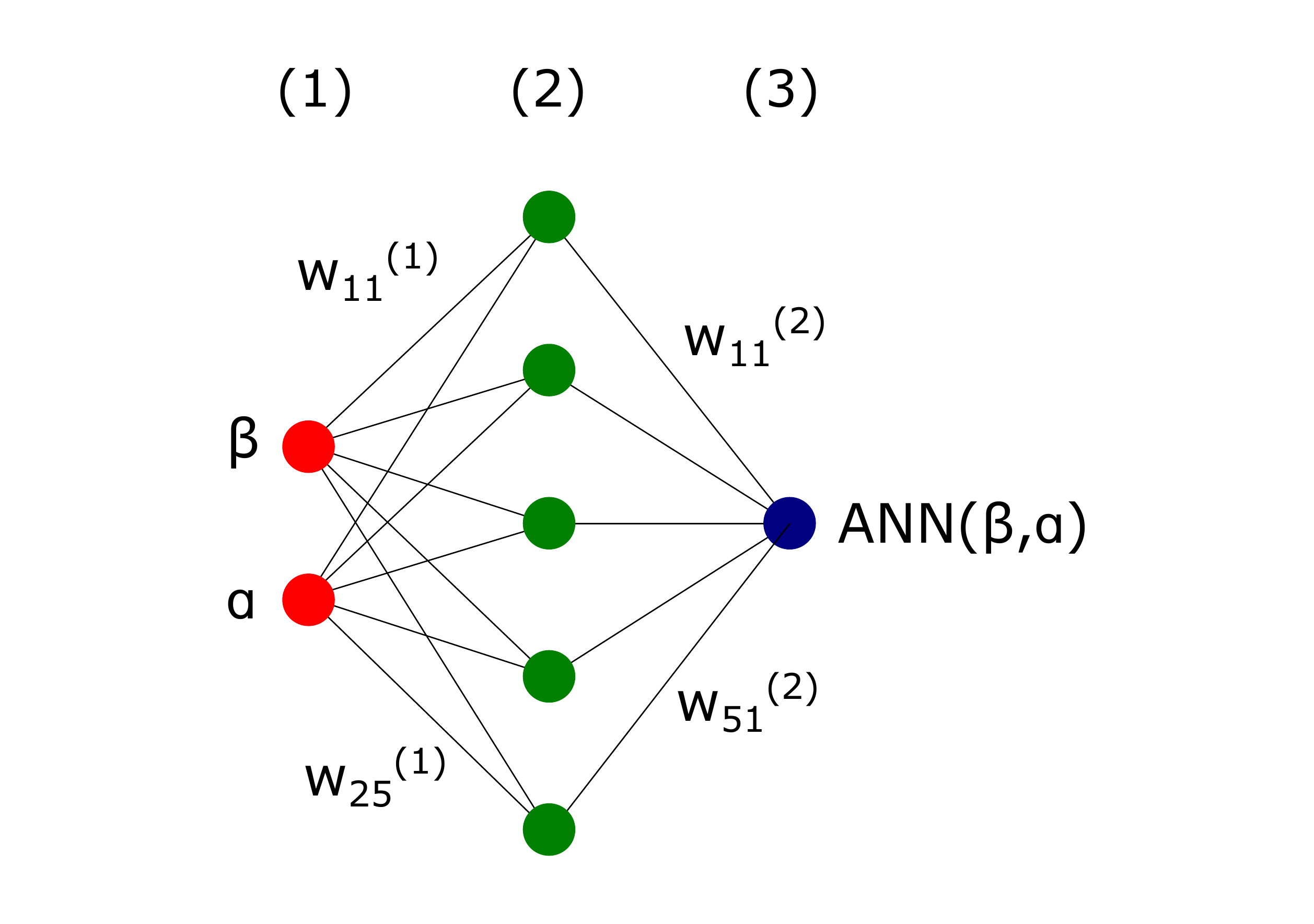}
\caption{Scheme of an exemplary artificial neural network used to represent the profile function in the double distribution model.}
\label{fig:baspace:ann:scheme}
\end{center}
\end{figure}

We use the following model for $h_{C}(\beta, \alpha)$ fulfilling all the aforementioned requirements:
\begin{equation}
h_{C}(\beta, \alpha) = \frac{\mathrm{ANN}_{C}(|\beta|, \alpha)}{\displaystyle\int_{-1+|\beta|}^{1-|\beta|} \mathrm{d}\alpha \mathrm{ANN}_{C}(|\beta|, \alpha) }  \,.
\label{eq:baspace:ann:ddann}
\end{equation}
Here, $\mathrm{ANN}_{C}(|\beta|, \alpha)$ is the output of a single artificial neural network. We have chosen a feed forward artificial neural network with a single hidden layer, whose example is shown in Fig.~\ref{fig:baspace:ann:scheme}. Our choice simplifies the evaluation of the integral in the denominator of Eq.~\eqref{eq:baspace:ann:ddann}. The signal is processed by the network in a way similar to the case described in Sect.~\ref{sec:xdep:ann}. Keeping the nomenclature consistent throughout the text, two neurons in the input layer distribute the $|\beta|$ and $\alpha$ values:  
\begin{flalign}
o_{1}^{(1)} \equiv &~ |\beta| \,, \\
o_{2}^{(1)} \equiv &~ \alpha \,,
\end{flalign}
while the neuron in the output layer gives:
\begin{flalign}
o_{1}^{(3)} \equiv &~ \mathrm{ANN}_{C}(|\beta|, \alpha) \,.
\end{flalign}
A number of neurons in the hidden layer process information according to this equation:
\begin{flalign}
o_{k}^{(2)}
& = \Big[ \varphi_{k}^{(2)}\left(b_{k}^{(2)} + w_{1,k}^{(1)}|\beta| + w_{2,k}^{(1)}\alpha/(1-|\beta|)\right)  \nonumber \\
& - \phantom{\Big[}\varphi_{k}^{(2)}\left(b_{k}^{(2)} + w_{1,k}^{(1)}|\beta| + w_{2,k}^{(1)}\right)\Big]  \nonumber \\
& + \Big[w_{2,k}^{(1)} \rightarrow -w_{2,k}^{(1)}\Big]\,,
\label{eq:baspace:ann:ddann_hidden}
\end{flalign}
where the biases, $b_{k}^{(2)}$, and weights, $w_{1,k}^{(1)}$, $w_{2,k}^{(1)}$, are free parameters of the network, and where $\varphi_{k}^{(2)}(\cdot)$ is the activation function to be fixed by the network architect. For instance, it can be the sigmoid function from Eq.~\eqref{eq:xxispace:act_sigmoid} or the ReLU function from Eq.~\eqref{eq:xxispace:act_relu}.
For the output layer we have:
\begin{flalign}
o_{1}^{(3)} = \sum_{k}  w_{k,1}^{(2)} o_{k}^{(2)} \,,
\end{flalign}
where we explicitly used the linear activation function. There is no bias parameter and the weights, $w_{k,1}^{(2)}$ are the other free parameters of the network. With the sigmoid function for $\varphi_{k}^{(2)}(\cdot)$, the normalisation factor is:
\begin{flalign}
\int_{-1+|\beta|}^{1-|\beta|} & \mathrm{d}\alpha \mathrm{ANN}_{C}(|\beta|, \alpha) = 
2 \sum_{k}
\frac{w_{k,1}^{(2)}}{w_{2,k}^{(1)}}\left(|\beta| - 1\right)
	\Big[ \nonumber \\
    & \phantom{-~}\log\left(\cosh\left(b_{k}^{(2)} - w_{2,k}^{(1)} + |\beta| w_{1,k}^{(1)}\right)\right) \nonumber \\
    & - \log\left(\cosh\left(b_{k}^{(2)} + w_{2,k}^{(1)} + |\beta| w_{1,k}^{(1)}\right)\right) \nonumber \\
    & + w_{2,k}^{(1)} \tanh\left(b_{k}^{(2)} - w_{2,k}^{(1)} + |\beta| w_{1,k}^{(1)}\right) \nonumber \\
    & + w_{2,k}^{(1)} \tanh\left(b_{k}^{(2)} + w_{2,k}^{(1)} + |\beta| w_{1,k}^{(1)}\right)
	\Big]
\end{flalign}

\subsection{$F_{S}(\beta, \alpha)$ contribution}
\label{sec:baspace:Fg}

The shadow contribution is modelled in a quite similar way to $F_{C}(\beta, \alpha)$:
\begin{equation}
   F_{S}(\beta, \alpha) = f(\beta) h_{S}(\beta, \alpha) \,.
\label{eq:baspace:ddfactorisationFg}
\end{equation}
Here, $f(\beta)$ has been already given in Eq.~\eqref{eq:baspace:pdf}. The inclusion of this factor may come as a surprise, as the shadow GPD vanishes in the forward limit, but $f(\beta)$ helps here to fulfil the positivity constraint. Similarly to $h_{C}(\beta, \alpha)$,  $h_{S}(\beta, \alpha)$ must be an even function with respect to both $\beta$ and $\alpha$ variables, and we make it to vanish at $|\alpha| + |\beta| = 1$. Its normalisation is however different. We require
\begin{equation}
\int_{-1+\beta}^{1-\beta} \mathrm{d}\alpha h_{S}(\beta, \alpha) = 0 \,,
\end{equation}
to ensure the vanishing at $\xi = 0$. We use the following model for $h_{S}(\beta, \alpha)$, which fulfils all the aforementioned requirements:
\begin{flalign}
h_{S}(\beta, \alpha) / N_{S} & = \frac{\mathrm{ANN}_{S}(|\beta|, \alpha)}{\displaystyle\int_{-1+|\beta|}^{1-|\beta|} \mathrm{d}\alpha \mathrm{ANN}_{S}(|\beta|, \alpha) } \nonumber \\
& - \frac{\mathrm{ANN}_{S'}(|\beta|, \alpha)}{\displaystyle\int_{-1+|\beta|}^{1-|\beta|} \mathrm{d}\alpha \mathrm{ANN}_{S'}(|\beta|, \alpha) }
\,.
\label{eq:baspace:ann:ddann_g}
\end{flalign}
Here, $N_{S}$ is a scaling parameter, while $\mathrm{ANN}_{S}(|\beta|, \alpha)$ and $\mathrm{ANN}_{S'}(|\beta|, \alpha)$ are neural networks. As one may conclude, effectively, we are constructing the shadow contribution by subtracting two GPDs having the same forward limit. One of these GPDs, related to $\mathrm{ANN}_{S}(|\beta|, \alpha)$, will be the subject of modelling. The other one, related to $\mathrm{ANN}_{S'}(|\beta|, \alpha)$, provides an arbitrary reference point. For simplicity we take:
\begin{equation}
\mathrm{ANN}_{S'}(|\beta|, \alpha) \equiv \mathrm{ANN}_{C}(|\beta|, \alpha) \,,
\end{equation}
that is, one of two networks used in $F_{S}(\beta, \alpha)$ is the same as that used in $F_{C}(\beta, \alpha)$. In fits presented in Sect.~\ref{sec:baspace:fits} $\mathrm{ANN}_{S}(|\beta|, \alpha)$ will have the same architecture as $\mathrm{ANN}_{S'}(|\beta|, \alpha)$, but different free parameters (weights and biases).

\subsection{$F_{D}(\beta, \alpha)$ contribution}
\label{sec:baspace:Fd}

This contribution gives the D-term:
\begin{equation}
F_{D}(\beta, \alpha) =  \delta(\beta) D(\alpha) \,,
\end{equation}
where: 
\begin{equation}
D(\alpha) = (1-\alpha^2) \sum_{\substack{i=1 \\ \mathrm{odd}}}^{N} d_{i}C_{i}^{\nicefrac{3}{2}}\left(\alpha\right) \,,
\label{eq:baspace:ann:Dterm}
\end{equation}
$d_{i}$ are the coefficients of the expansion on the family of Gegenbauer polynomials and $N$ is a truncation parameter. This definition in $(x,\xi)$-space explicitly gives:
\begin{equation}
D(z) = \Theta(1-|z|)(1-z^2) \sum_{\substack{i=1 \\ \mathrm{odd}}}^{N} d_i C_{i}^{\nicefrac{3}{2}}\left(z\right) \,,
\end{equation}
where $z=x/\xi$.

\subsection{Remarks}

In Eq.~\eqref{eq:baspace:ann:ddann_hidden} the variable $\alpha$ is scaled by $1-|\beta|$, such that $\alpha/(1-|\beta|)$ spans over the range of $(-1, 1)$. Such a standardisation of variables is typical for neural networks, as it leads to a faster convergence and it allows to describe equally well the dependencies on all input variables. 

The factorisation expressed by Eq.~\eqref{eq:baspace:ddfactorisationFu} is arbitrary. In principle, the neural network could absorb some, if not all, information about PDFs. For instance, we could express a double distribution $F$ in the following way:
\begin{equation}
F(\beta, \alpha) = \mathrm{sgn}(\beta)|\beta|^{-\delta}(1-|\beta|)^{\gamma}\mathrm{ANN}(\beta, \alpha) \,,
\end{equation}
where $\delta$ and $\gamma$ are powers driving the behaviour of PDF at $x \rightarrow 0$ and $x \rightarrow 1$, respectively, and where the normalisation factor, as used in the denominator of Eq.~\eqref{eq:baspace:ann:ddann}, is not needed anymore. This kind of factorisation may be useful when (semi-)inclusive and exclusive data will be simultaneously used to constrain GPDs.

%% file: sec_ba_fits.tex
\subsection{Fits to data}
\label{sec:baspace:fits}

We now discuss technicalities of constraining GPD models defined in $(\beta, \alpha)$-space from data. The examples presented in this section demonstrate possible applications of our approach, but also its performance in conditions imposed by currently available data. We start with the presentation of results, and then we elaborate more on aspects of our numerical analysis like minimisation, propagation of uncertainties, treatment of outliers, regularisation and positivity enforcement. 

We perform three extractions of GPDs with pseudo-data generated with the GK model \cite{Goloskokov:2005sd,Goloskokov:2007nt,Goloskokov:2009ia}. In all three cases $\mathrm{ANN}_{C}(|\beta|, \alpha)$ used in both $F_{C}(\beta, \alpha)$ and $F_{S}(\beta, \alpha)$, and $\mathrm{ANN}_{S}(|\beta|, \alpha)$ used in $F_{S}(\beta, \alpha)$ consist of $5$ neurons each, and we use the sigmoid activation function.  The $F_{D}(\beta, \alpha)$ contribution consists of $5$ elements, \ie the sum in Eq.~\eqref{eq:baspace:ann:Dterm} runs up to $i = 9$. 

\begin{figure}[!ht]
\begin{center}
\includegraphics[width=0.39\textwidth]{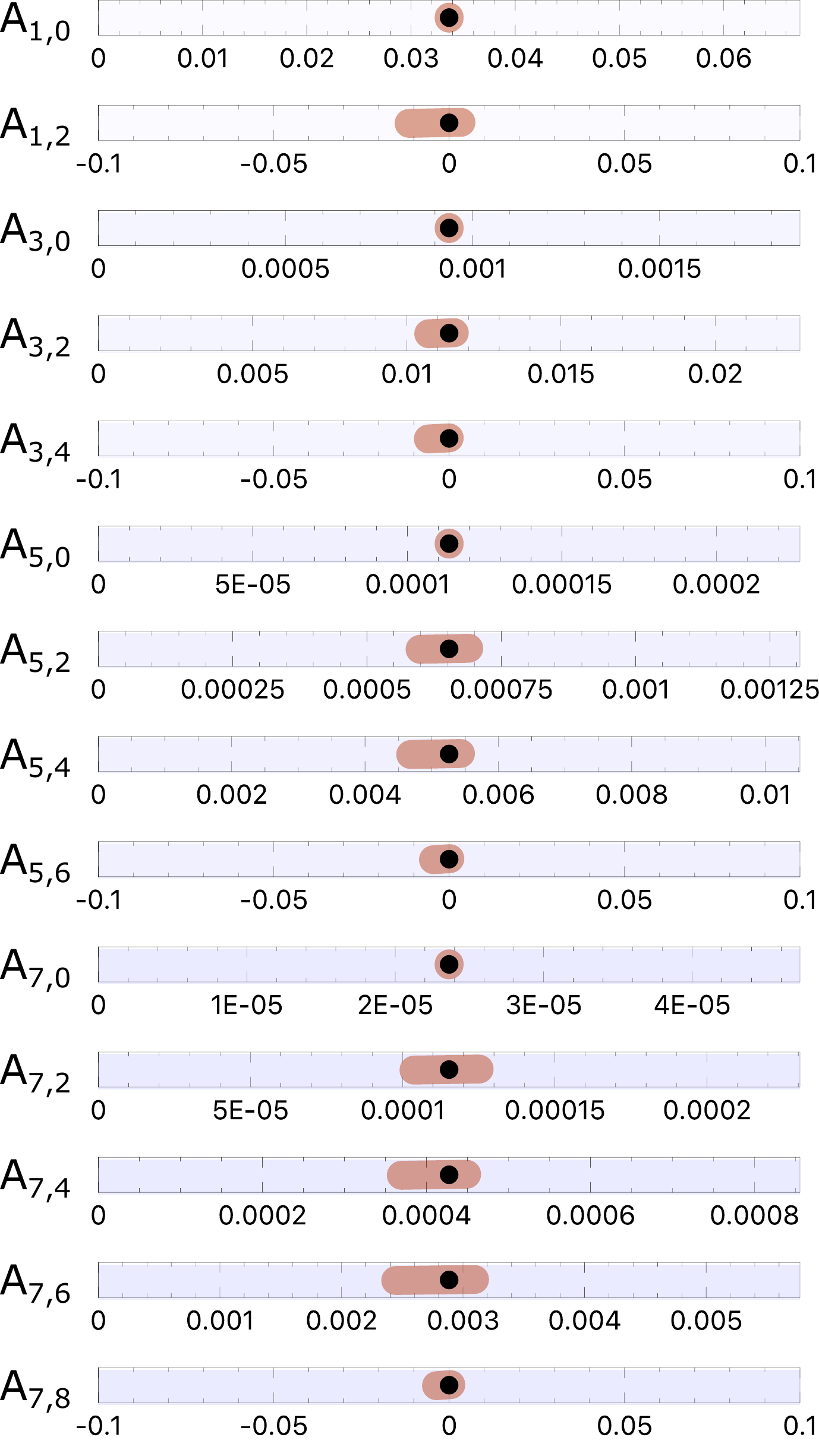}
\caption{Result of constraining the ANN model using $400$ points evaluated with the GK model \cite{Goloskokov:2005sd,Goloskokov:2007nt,Goloskokov:2009ia} for sea quarks: $x \neq \xi$ case, positivity not enforced, all three contributions ($F_{C}(\beta, \alpha), F_{S}(\beta, \alpha), F_{D}(\beta, \alpha)$) used in the presentation of results. The plot shows the first Mellin coefficient for odd moments (even moments strictly vanish) evaluated for the GK (black dots) and ANN (solid bands) models.}
\label{fig:fits:moments}
\end{center}
\end{figure}

\begin{figure*}[!ht]
\begin{center}
\includegraphics[width=0.30\textwidth]{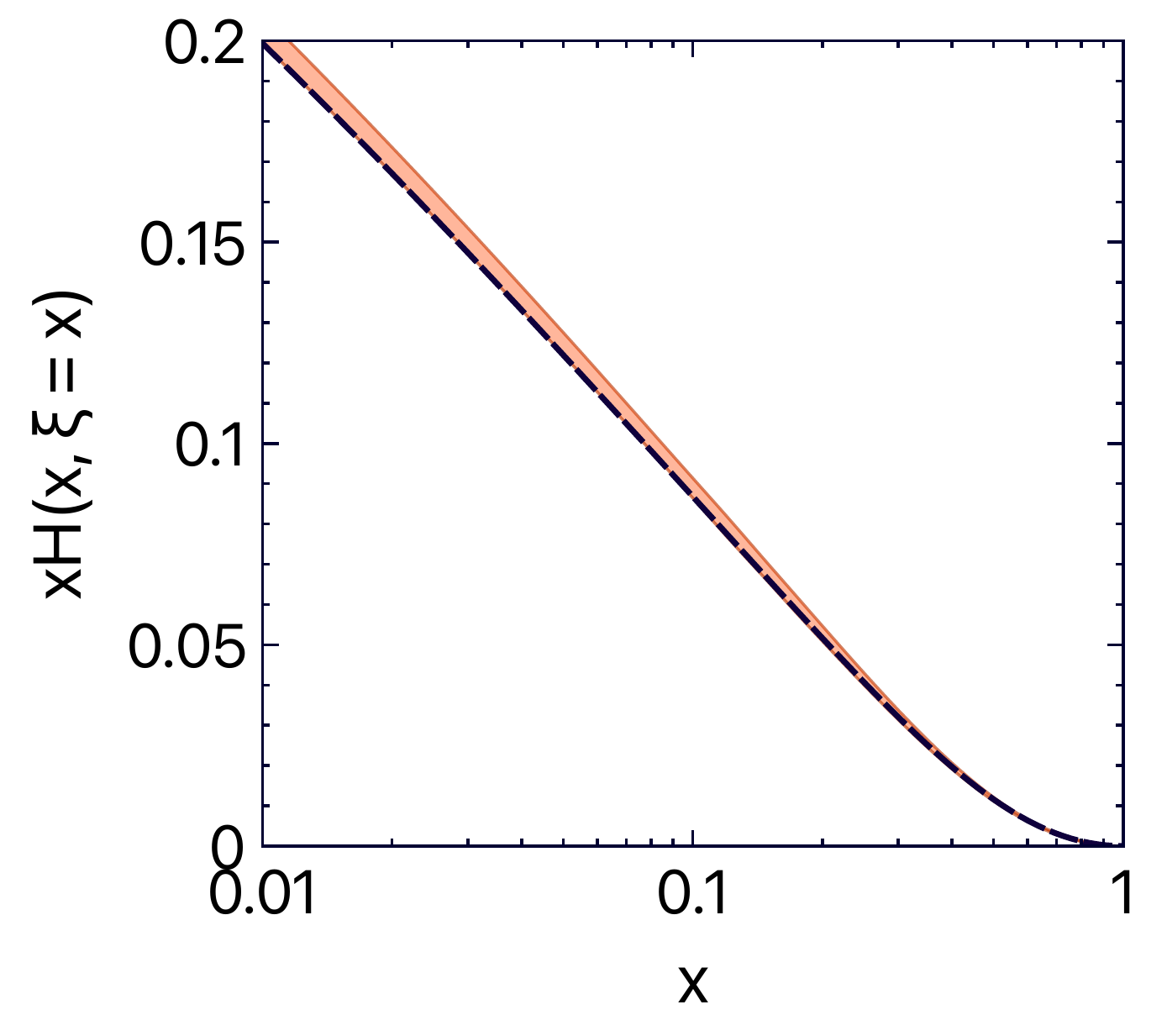}
\includegraphics[width=0.30\textwidth]{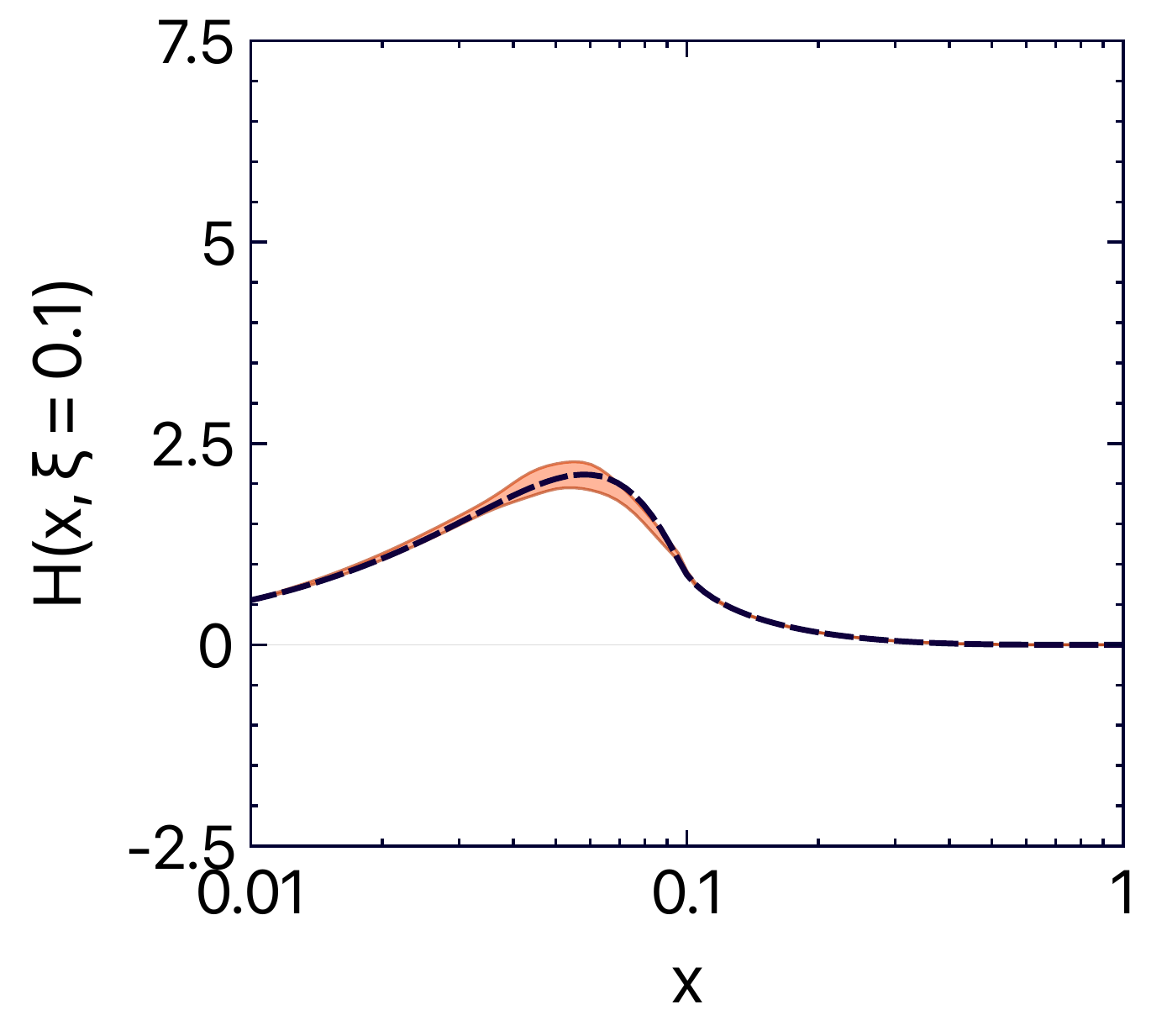}
\includegraphics[width=0.30\textwidth]{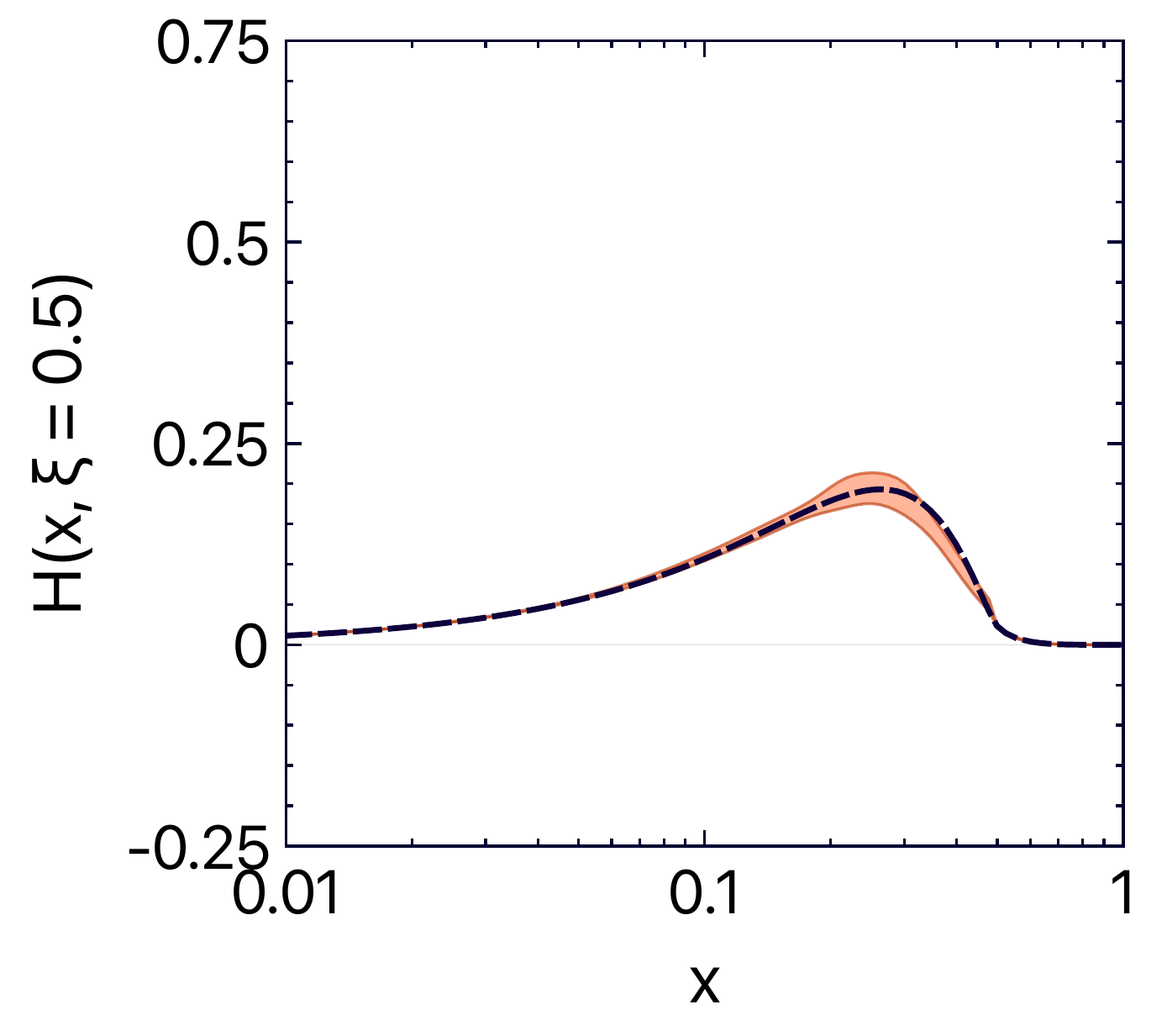}
\caption{Result of constraining the ANN model using $400$ points evaluated with the GK model \cite{Goloskokov:2005sd,Goloskokov:2007nt,Goloskokov:2009ia} for sea quarks: $x \neq \xi$ case, positivity not enforced, all three contributions ($F_{C}(\beta, \alpha), F_{S}(\beta, \alpha), F_{D}(\beta, \alpha)$) shown. See the text for more details. The comparison is for (left) $\xi = x$, (center) $\xi = 0.1$ and (right) $\xi = 0.5$. The dashed lines denote the GK model, while the bands represent the result of the fit in the form of a $68\%$ confidence level.}
\label{fig:fits:xNeqxi}
\end{center}
\end{figure*}

In the first test-case we train our model on $N_{\mathrm{pts}} = 400$ points uniformly covering the domain of $-4 < \log_{10}(x) < \log_{10}(0.95)$, $-4 < \log_{10}(\xi) < \log_{10}(0.95)$, $t=0$ and $Q^2 = 4~\mathrm{GeV}^2$. That is, in this scenario, $x \neq \xi$ data are used to constrain our ANN model. The purpose of this test is to check if our approach can be used to reproduce such GPD models as GK. All three contributions from Eq.~\eqref{eq:baspace:principles:mainAnsatz} are used in the fit, even if we \emph{a priori} know that GK does not include any D-term. It is because $F_{C}(\beta, \alpha)$ and $F_{S}(\beta, \alpha)$ may contribute to the D-term due to the $(1-x^2)$ and $(x^2-\xi^2)$ factors, respectively. With $F_{D}(\beta, \alpha)$ included in the fit we are able to compensate for those contributions. We do not enforce the positivity constraint, as the GK model violates the inequality shown in Eq.~\eqref{eq:theory:positivity}. Enforcing this positivity inequality would lead to a discrepancy between our model and GK. The result is shown in  Fig.~\ref{fig:fits:moments} for Mellin moments and in Fig.~\ref{fig:fits:xNeqxi} for the $x$-dependence. We observe a very good agreement between fitted ANN model and GK, hence we conclude to a successful outcome of this test.

\begin{figure*}[!ht]
\begin{center}
\includegraphics[width=0.30\textwidth]{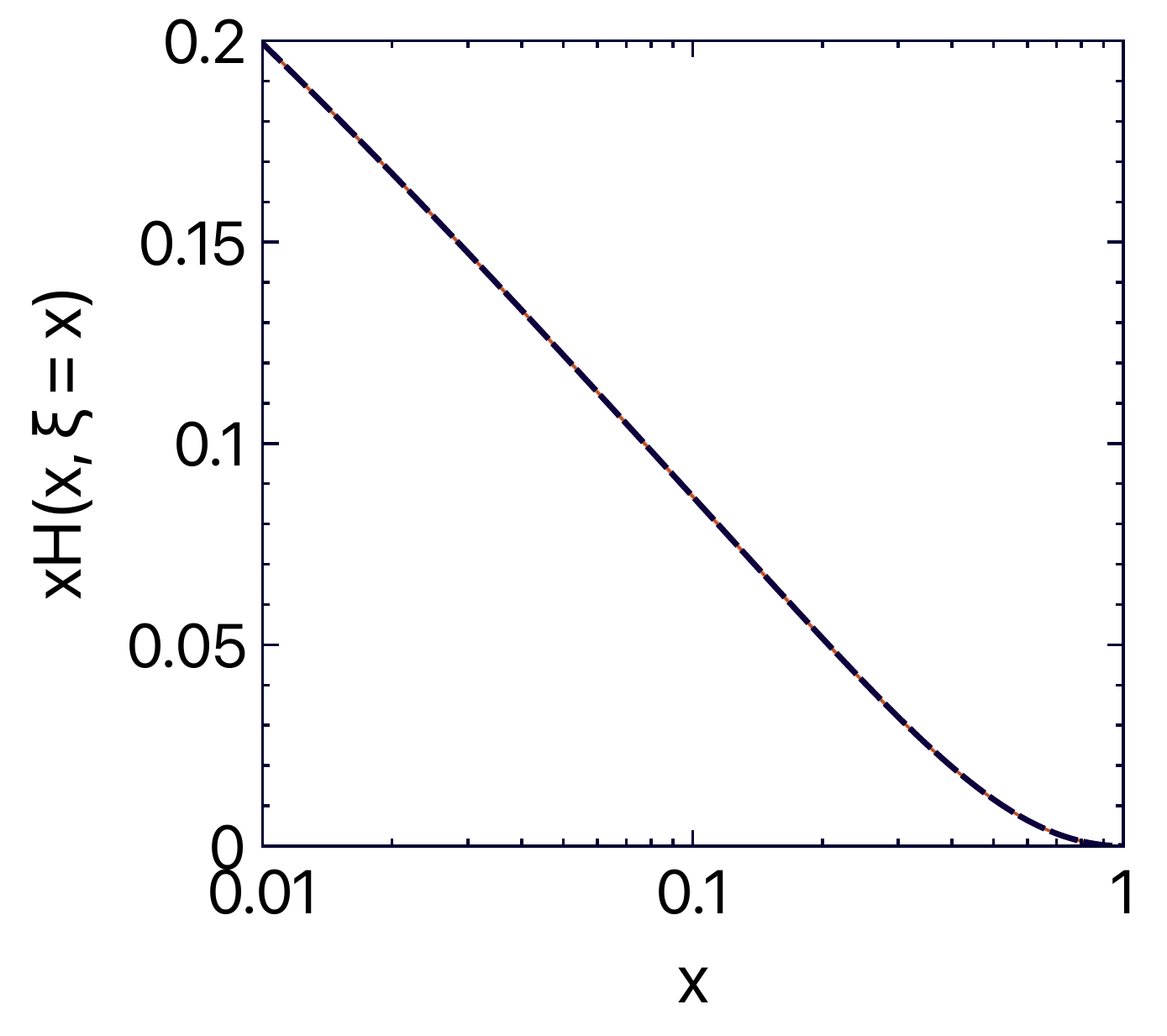}
\includegraphics[width=0.30\textwidth]{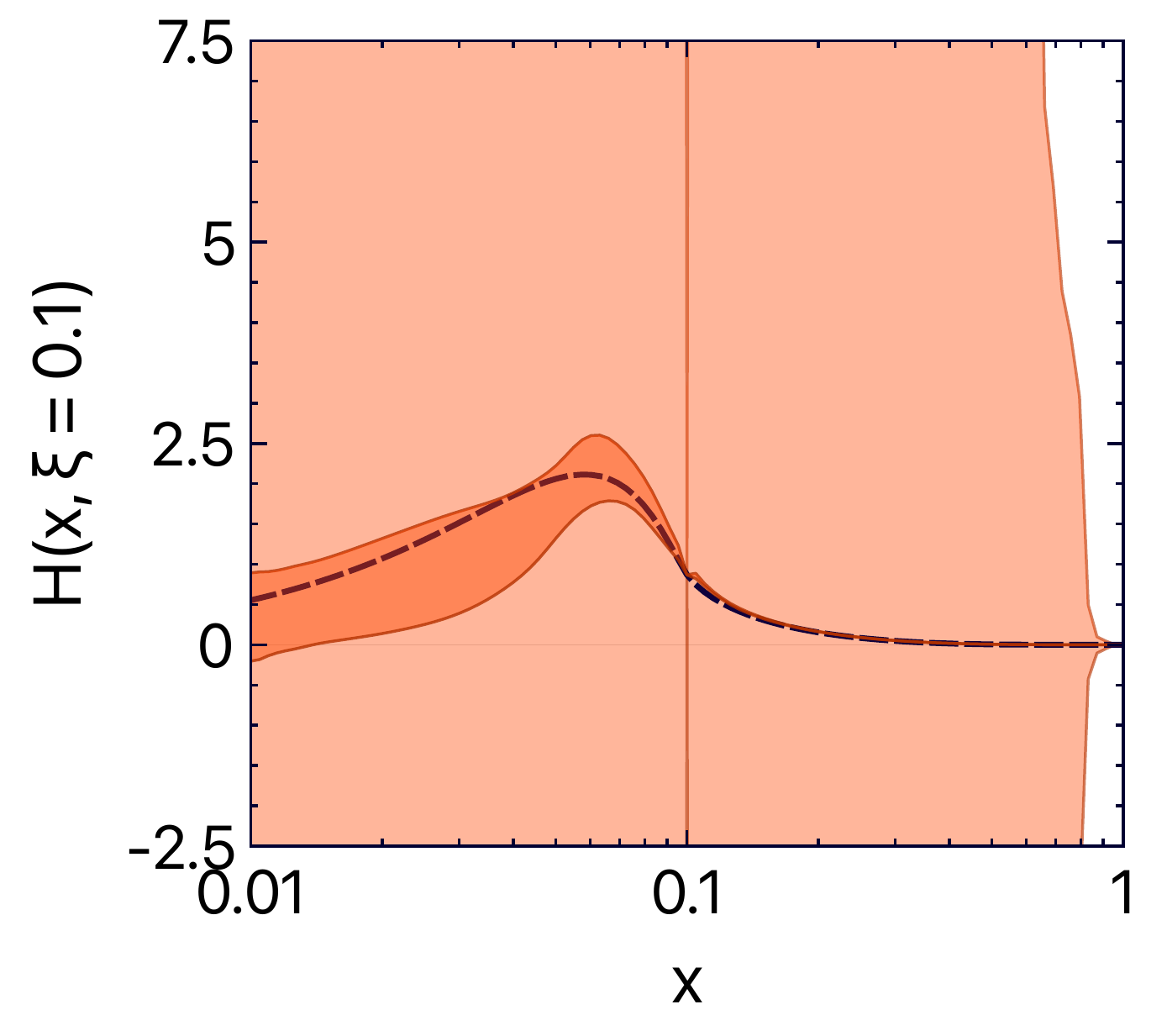}
\includegraphics[width=0.30\textwidth]{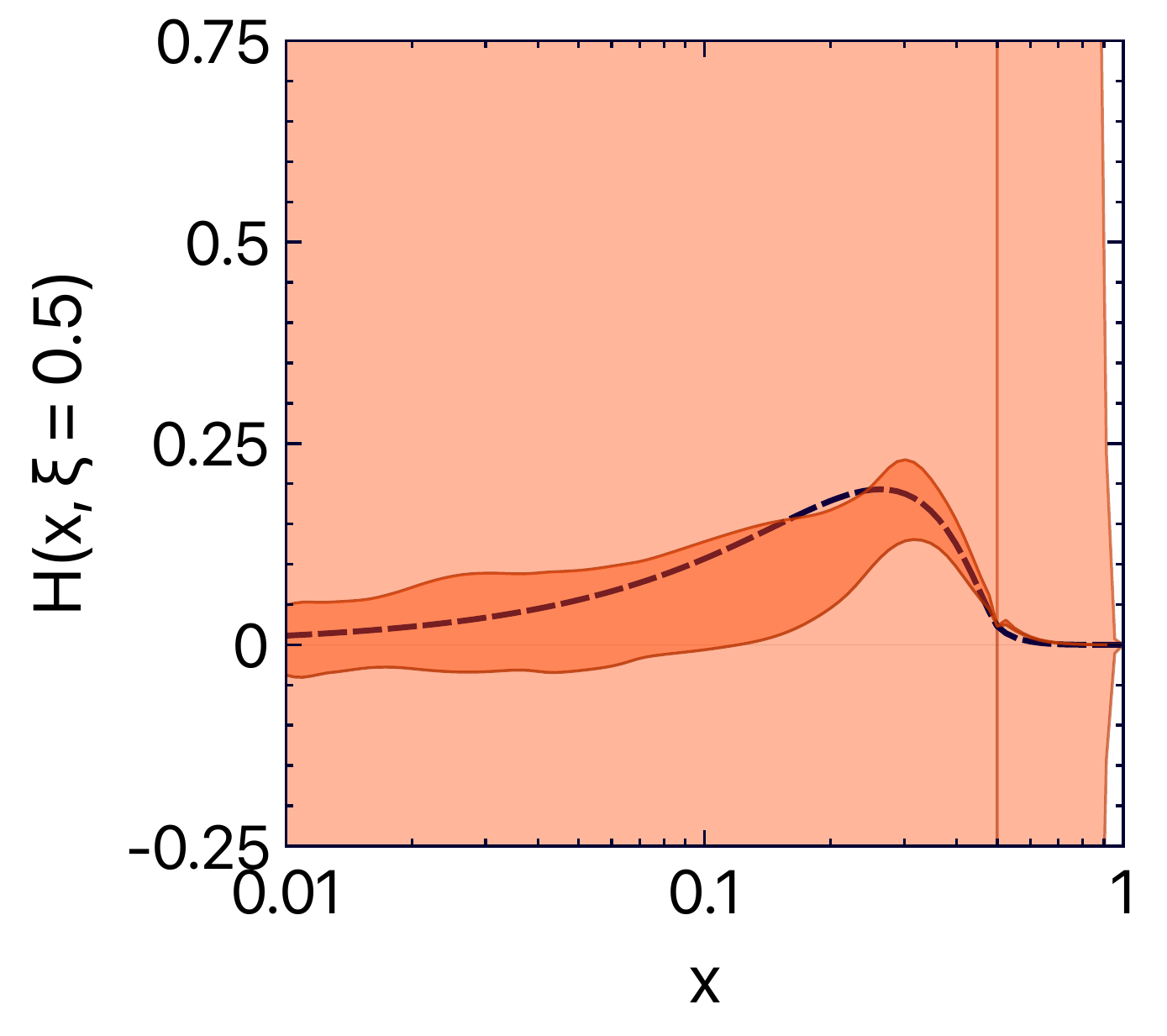}
\caption{Result of constraining the ANN model using $200$ points evaluated with GK model \cite{Goloskokov:2005sd,Goloskokov:2007nt,Goloskokov:2009ia} for sea quarks: $x = \xi$ case, positivity not enforced, $F_{D}(\beta, \alpha)$ not shown. See the text for more details. The comparison is for (left) $x = \xi$, (center) $\xi = 0.1$ and (right) $\xi = 0.5$. The dashed lines denote the GK model, while the bands represent the result of the fit in the form of a $68\%$ confidence level. The inner bands show $F_{C}(\beta, \alpha)$ contribution, alone, while the outer bands are for $F_{C}(\beta, \alpha) + F_{S}(\beta, \alpha)$.}
\label{fig:fits:xEqxiNoPos}
\end{center}
\end{figure*}

In the second test-case only $x = \xi$ data are used to constrain our GPD model. These are $N_{\mathrm{pts}} = 200$ points uniformly covering the domain of $-4 < \log_{10}(x = \xi) < \log_{10}(0.95)$, $t=0$ and $Q^2 = 4~\mathrm{GeV}^2$. The purpose of this test is to demonstrate how our approach can be used to reconstruct GPDs from amplitudes for processes like DVCS and TCS, when described at LO. The positivity inequality is not enforced. The result is shown in Fig.~\ref{fig:fits:xEqxiNoPos}. Since we are not interested in the D-term (it is not constrained by $x=\xi$ data, but can be accessed elsewhere, namely in dispersive analyses) the contribution coming from $F_{D}(\beta, \alpha)$ is removed from the presentation of results. The contribution to the D-term generated by $F_{C}(\beta, \alpha)$ and $F_{S}(\beta, \alpha)$ is small with respect to the uncertainties and will be discussed in more detail in the next paragraph. In Fig.~\ref{fig:fits:xEqxiNoPos} one may observe exploding uncertainties, except for the $\xi = x$ line. This behaviour is expected. From the same figure one may judge on the importance of including the shadow contribution, $F_{S}(\beta, \alpha)$. We see that $F_{C}(\beta, \alpha)$ is over-constrained by the necessity of reproducing the GPD at both $\xi=0$ and $\xi = x$ lines. Since $h_{C}(\beta,\alpha)$ is normalized, see Eqs. \eqref{eq:ba:normTo1} and \eqref{eq:baspace:ann:ddann}, the neural network lacks the flexibility allowing for a significant contribution to the uncertainties in kinematic domains that are unconstrained by data.

\begin{figure*}[!ht]
\begin{center}
\includegraphics[width=0.30\textwidth]{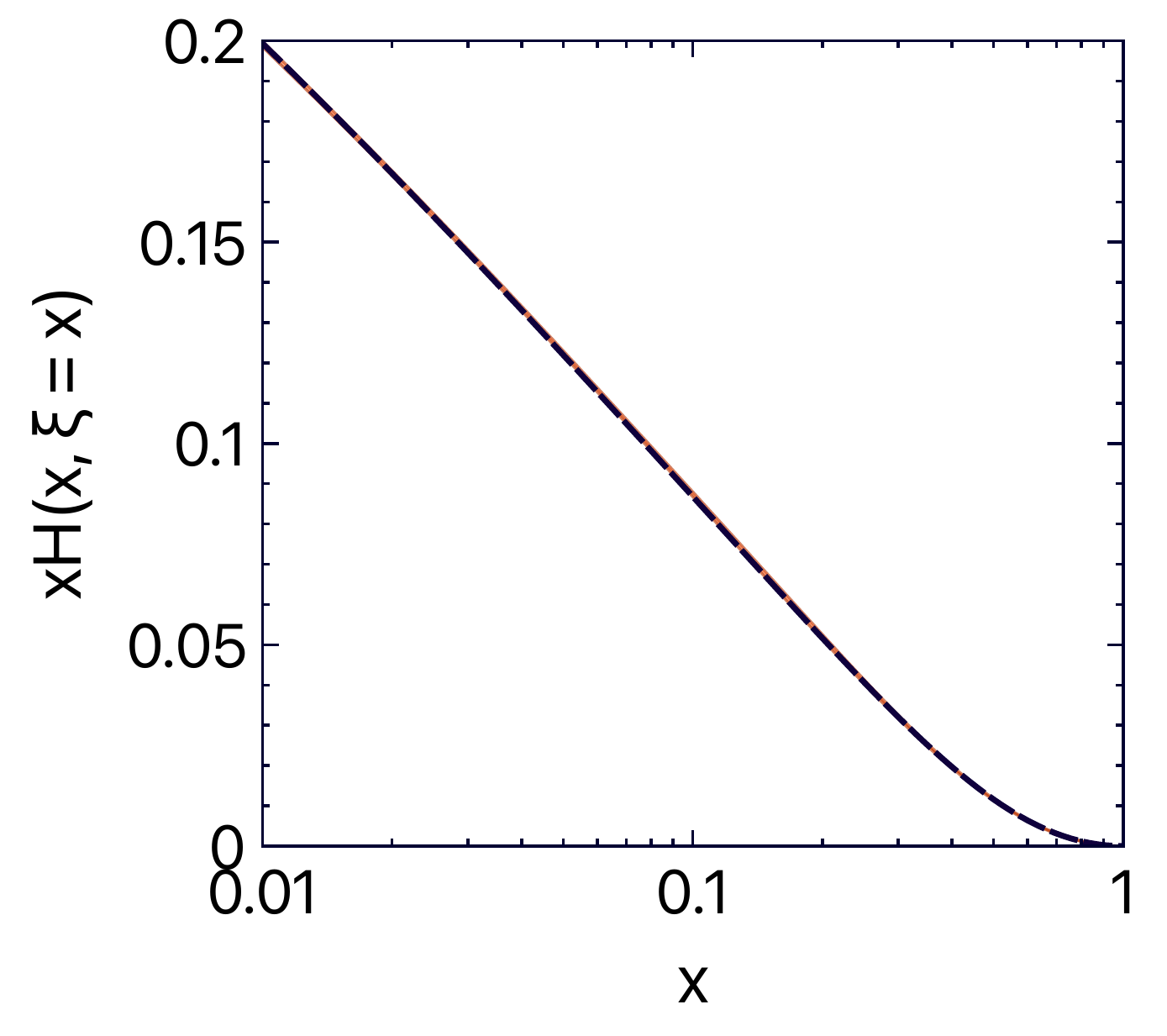}
\includegraphics[width=0.30\textwidth]{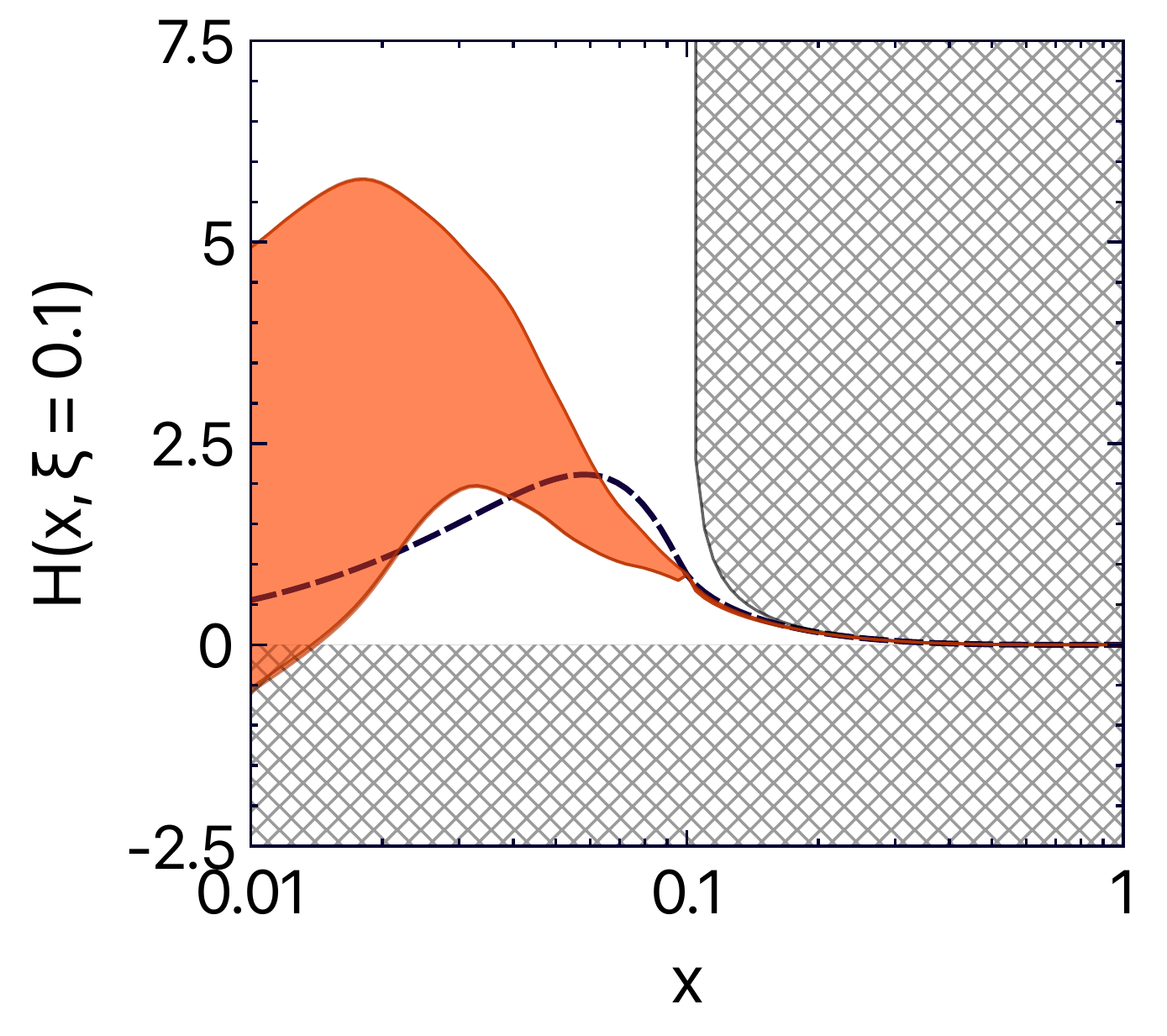}
\includegraphics[width=0.30\textwidth]{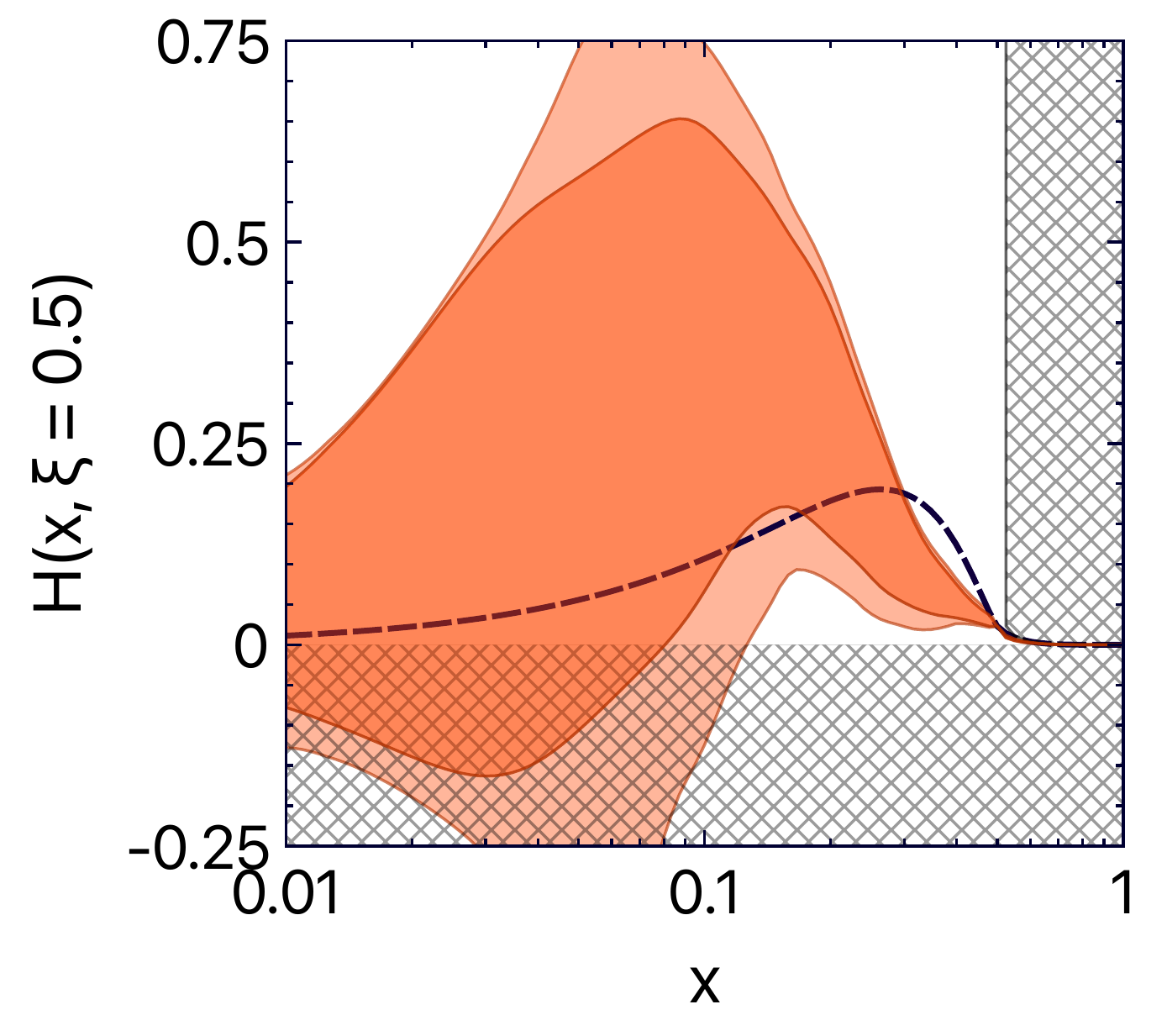}
\caption{Result of constraining the ANN model using $200$ points evaluated with the GK model \cite{Goloskokov:2005sd,Goloskokov:2007nt,Goloskokov:2009ia} for sea quarks: $x = \xi$ case, positivity enforced, $F_{D}(\beta, \alpha)$ not shown. See the text for more details. The comparison is for (left) $x = \xi$, (center) $\xi = 0.1$ and (right) $\xi = 0.5$. The dashed lines denote the GK model, while the bands represent the result of the fit in the form of a $68\%$ confidence level. The inner bands show  $F_{C}(\beta, \alpha)$ contribution, alone, while the outer bands are for $F_{C}(\beta, \alpha) + F_{S}(\beta, \alpha)$. The regions excluded by the positivity constraint are denoted by the hatched bands.}
\label{fig:fits:xEqxiPos}
\end{center}
\end{figure*}

\begin{figure*}[!ht]
\begin{center}
\includegraphics[width=0.30\textwidth]{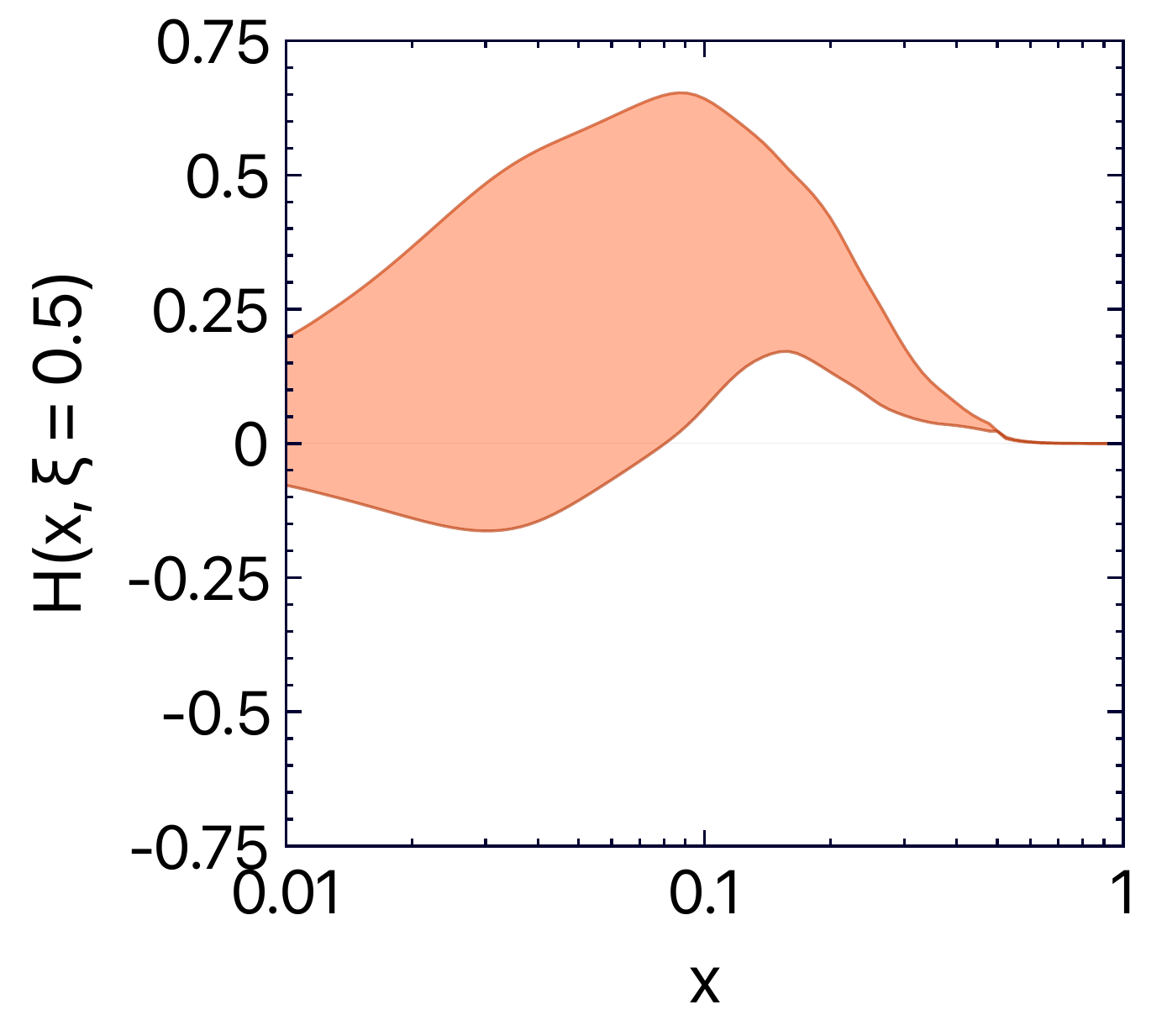}
\includegraphics[width=0.30\textwidth]{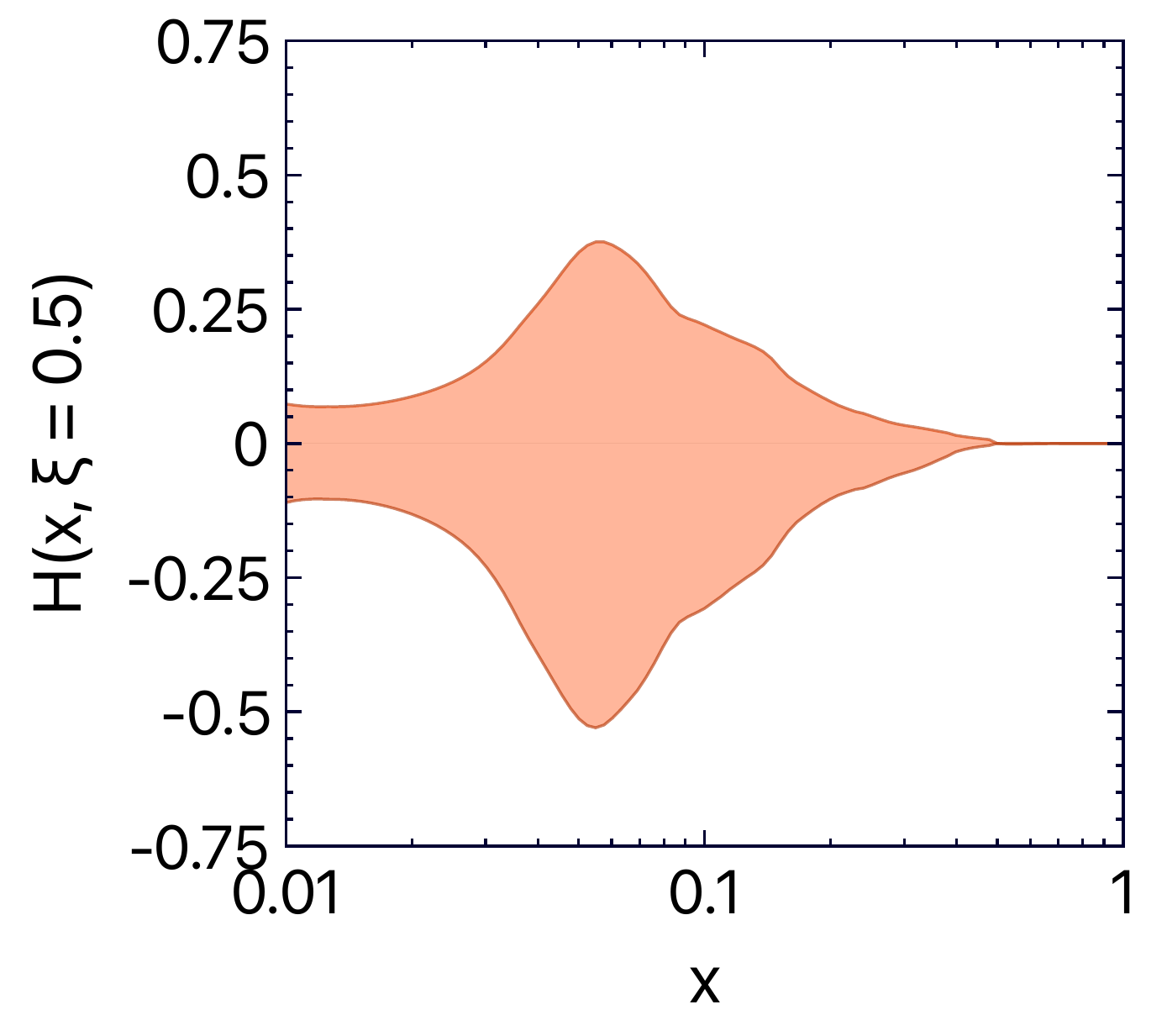}
\includegraphics[width=0.30\textwidth]{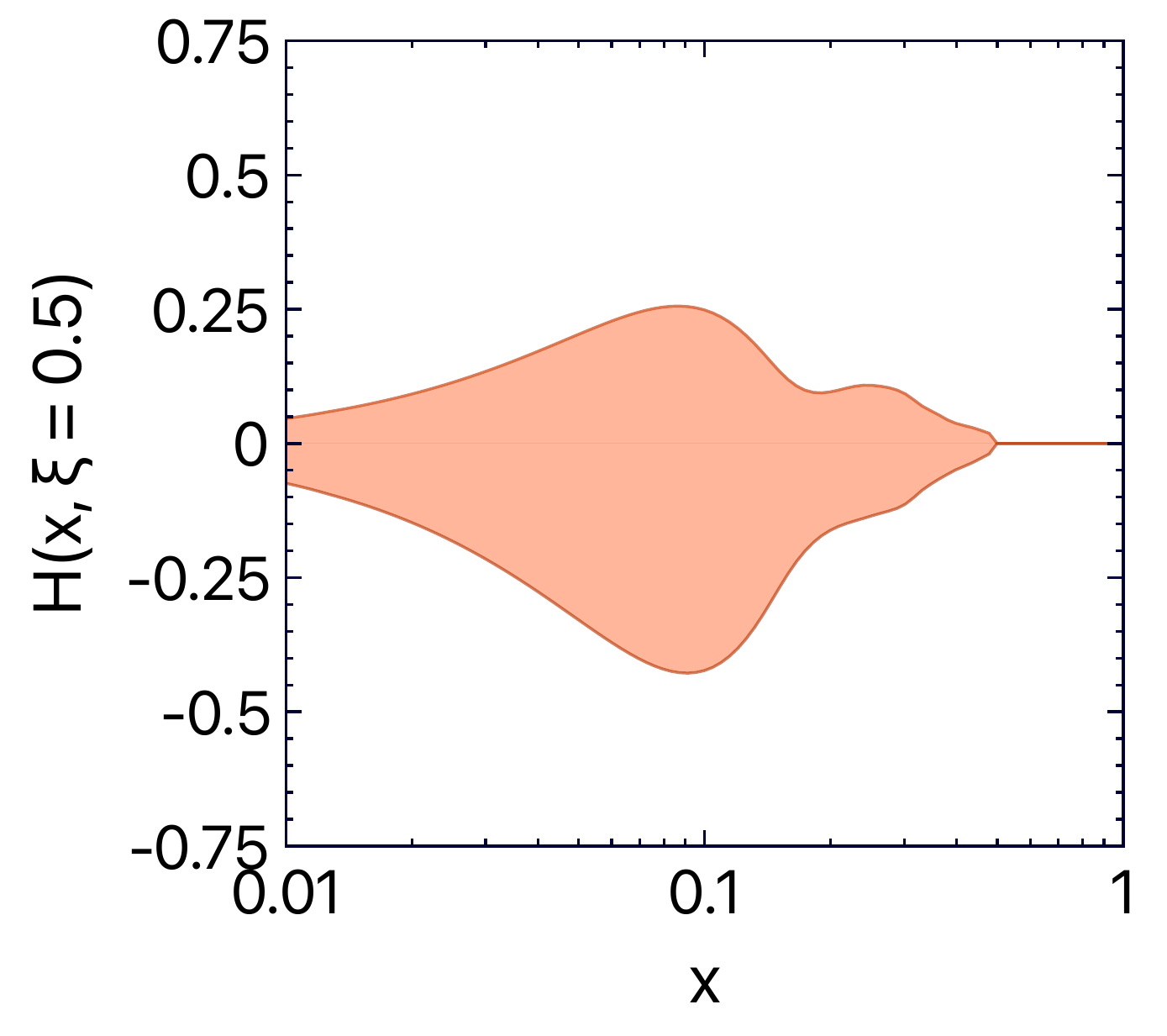}
\caption{Result of constraining the ANN model using $200$ points evaluated with the GK model \cite{Goloskokov:2005sd,Goloskokov:2007nt,Goloskokov:2009ia} for sea quarks: $x = \xi$ case, positivity enforced.  Contributions coming from (left) $F_{C}(\beta, \alpha)$ and (center) $F_{S}(\beta, \alpha)$ are shown. (right) Contribution to D-term induced by both $F_{C}(\beta, \alpha)$ and $F_{S}(\beta, \alpha)$ is shown. All plots are for $\xi=0.5$.}
\label{fig:fits:xEqxiPosContributions}
\end{center}
\end{figure*}

The conditions used in the third test-case are the same as in the second one, except now the positivity inequality is enforced to show its impact on the reduction of uncertainties. The result is shown in Fig.~\ref{fig:fits:xEqxiPos}. We observe a significant reduction of uncertainties due to positivity. The contributions coming from the $F_{C}(\beta, \alpha)$ and $F_{S}(\beta, \alpha)$ terms, together with the D-term induced by those two terms, is shown in Fig.~\ref{fig:fits:xEqxiPosContributions}. Both contributions are substantial.

In this analysis the minimisation, \ie the procedure of constraining free parameters, is done  with a genetic algorithm \cite{Mitchell:1998:IGA:522098}. This algorithm is iterative. In a single iteration many sets of free parameters, referred to in the literature as ``candidates'', are simultaneously checked against a ``fitness function''. After evaluation, the best candidates, \ie those characterised by the best values of the fitness function, are ``crossed over'' with a hope of obtaining even better candidates to be used in the next iteration. The cross-over is followed by a ``mutation'', where a number of free parameters is randomly changed, allowing for a significant reduction of the risk of converging to a local minimum. We note that, since in a given iteration of the genetic algorithm the fitness function is simultaneously evaluated for all candidates, multithreading computing can be employed to improve the performance of the minimisation. 

The employed fitness function is the root mean squared relative error \cite{GOCKEN2016320}:
\begin{equation}
\mathrm{RMSRE} = \sqrt{\frac{1}{N_{\mathrm{pts}}}\sum_{i=1}^{N_{\mathrm{pts}}} \left(\frac{H_{0}(x_{i},\xi_{i}) - H(x_{i},\xi_{i})}{H_{0}(x_{i},\xi_{i})}\right)^{2}} \,.
\end{equation}
Here, the sum runs over $N_{\mathrm{pts}}$ points probing $(x,\xi)$ phase-space, while $H_{0}(x_{i},\xi_{i})$ and $H(x_{i},\xi_{i})$ are GPDs given by the reference (here GK) and ANN models, respectively. The RMRSE allows to avoid significant differences of contributions to the fitness function coming from various domains of $(x,\xi)$. 

The pseudo-data used in this analysis do not have uncertainties. However, we are still interested to know what are the uncertainties of models in domains unconstrained by data, \ie what is the effect of the sparsity of data, in particular when only $x=\xi$ data are used in the fit and the neural network was designed to explore the whole $x\neq\xi$ domain. To estimate this kind of uncertainties we repeat the minimization multiple times, each time starting the genetic algorithms with a random set of initial parameters. We refer to the outcome of such a single minimization as ``a replica''. A replica is a possible realisation of the GPD model. All replicas reproduce the data used in the minimization, but because of the flexibility of ANNs their behaviour in unconstrained kinematic domains can be considered random. We note that this randomness can be unintentionally suppressed if one uses too small networks or restricts too much the values of weights and biases. 

We use the spread of $100$ replicas to estimate the model uncertainty in a given kinematic point. Because of the complexity of our fits and unavoidable problems in the minimisation, like falling into local minima, it sometimes happens that a single replica gives values that are very different compared to those given by other replicas. This problem is typical in phenomenological studies (see \eg Ref.~\cite{Ball:2012cx}), and requires finding and removing the problematic replicas, or suppressing them in the estimation of uncertainties. In this analysis we employ the following algorithm: for a given population of replicas \emph{i}) we randomly choose $1000$ points of $(x, \xi)$, \emph{ii}) for a given point we evaluate the GPD values that the replicas give in that point, and then we check which of those values are outliers by applying the $3\sigma$-rule \cite{IlyasC19}, \emph{iii}) the replicas giving values identified as outliers in more than $10\%$ points are entirely removed from the estimation of uncertainties. 

As in other analyses using effectively nonparametric models, a regularisation must be used to avoid biased results caused by over-fitting. Without a regularisation, an ANN tends to describe the training data extremely precisely, resulting in a minimal variance evaluated for these data. It does not mean though that the ANN will describe equally well any other data, \ie that it will have a predictive power. In general, a bias may appear, because too much attention was paid to describe the  training data, and the ANN does not describe well the general trends. Many types of regularisation methods exist, and the selection of a given method typically depends on the problem being under consideration. In this analysis we use the so-called dropout method \cite{JMLR:v15:srivastava14a}. In this method in a given iteration of the minimisation algorithm (here: the genetic algorithm) a predefined fraction of neurons (here: $10\%$) is randomly dropped, \ie  some neurons become inactive, they do not process the signal. The output of other neurons is correspondingly scaled to compensate the loss. Effectively, in each iteration a different architecture of ANN is probed, avoiding focusing on details only characterising the training sample. 

\begin{figure*}[!ht]
\begin{center}
\includegraphics[width=0.35\textwidth]{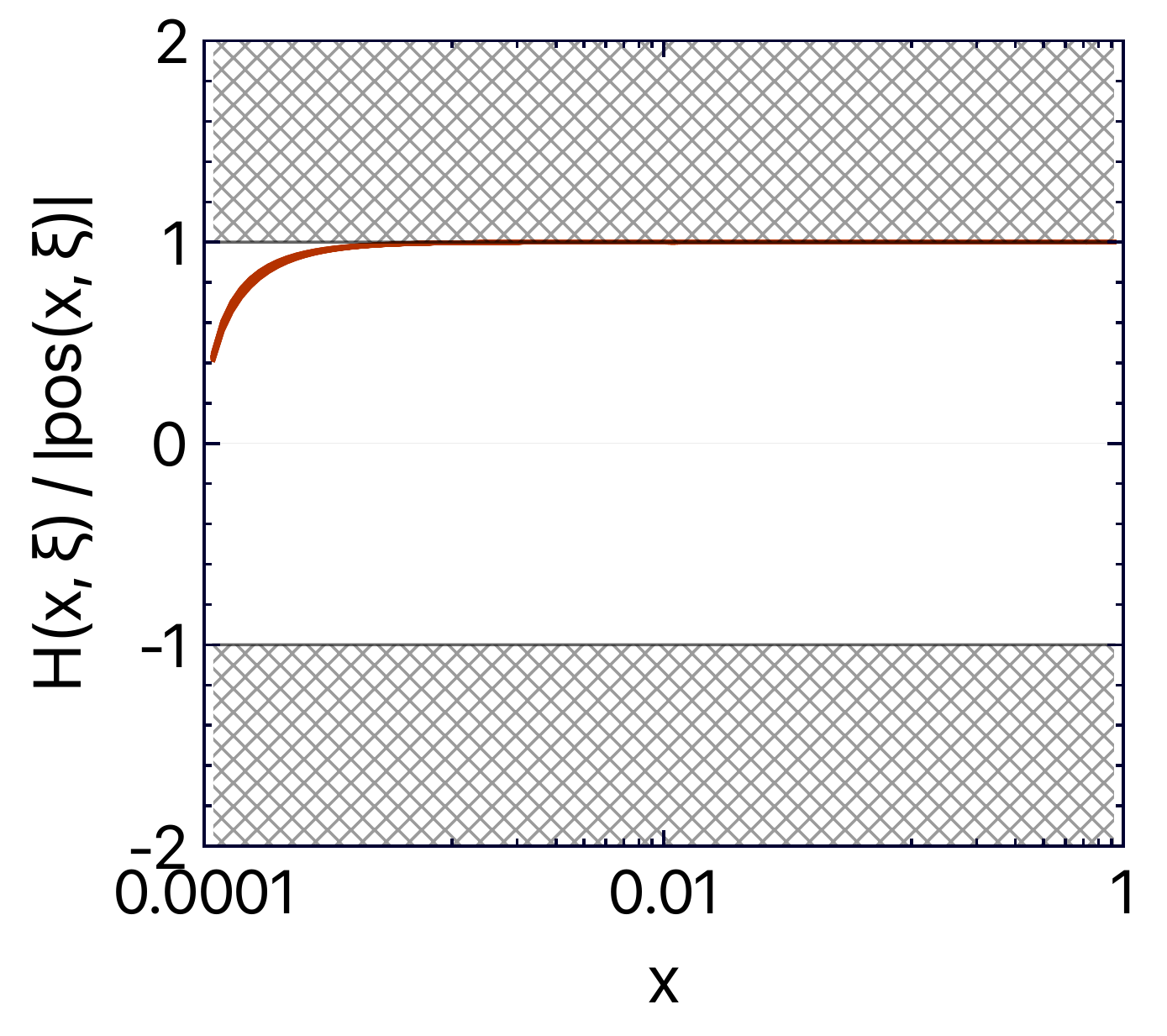}
\includegraphics[width=0.35\textwidth]{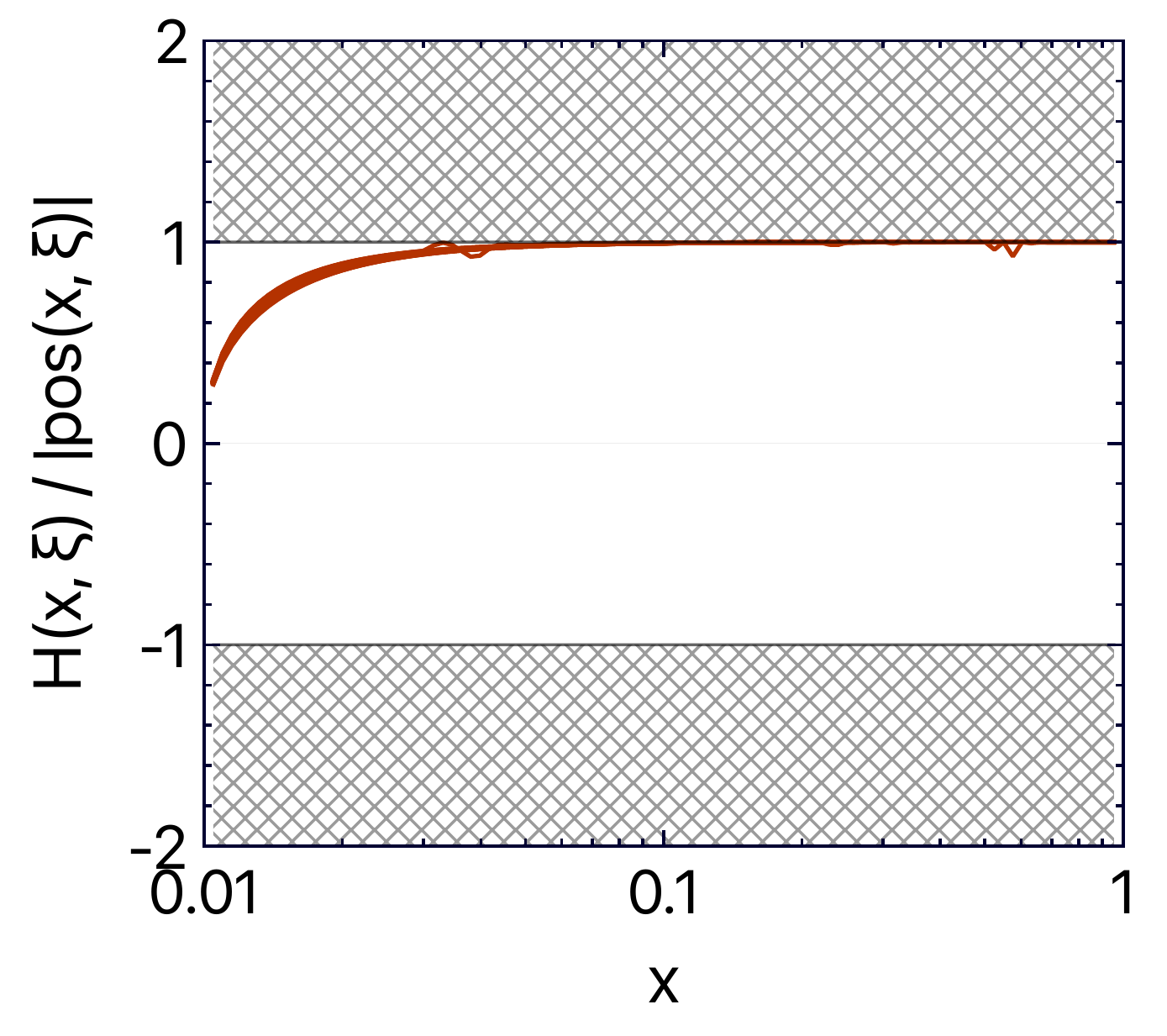}
\includegraphics[width=0.35\textwidth]{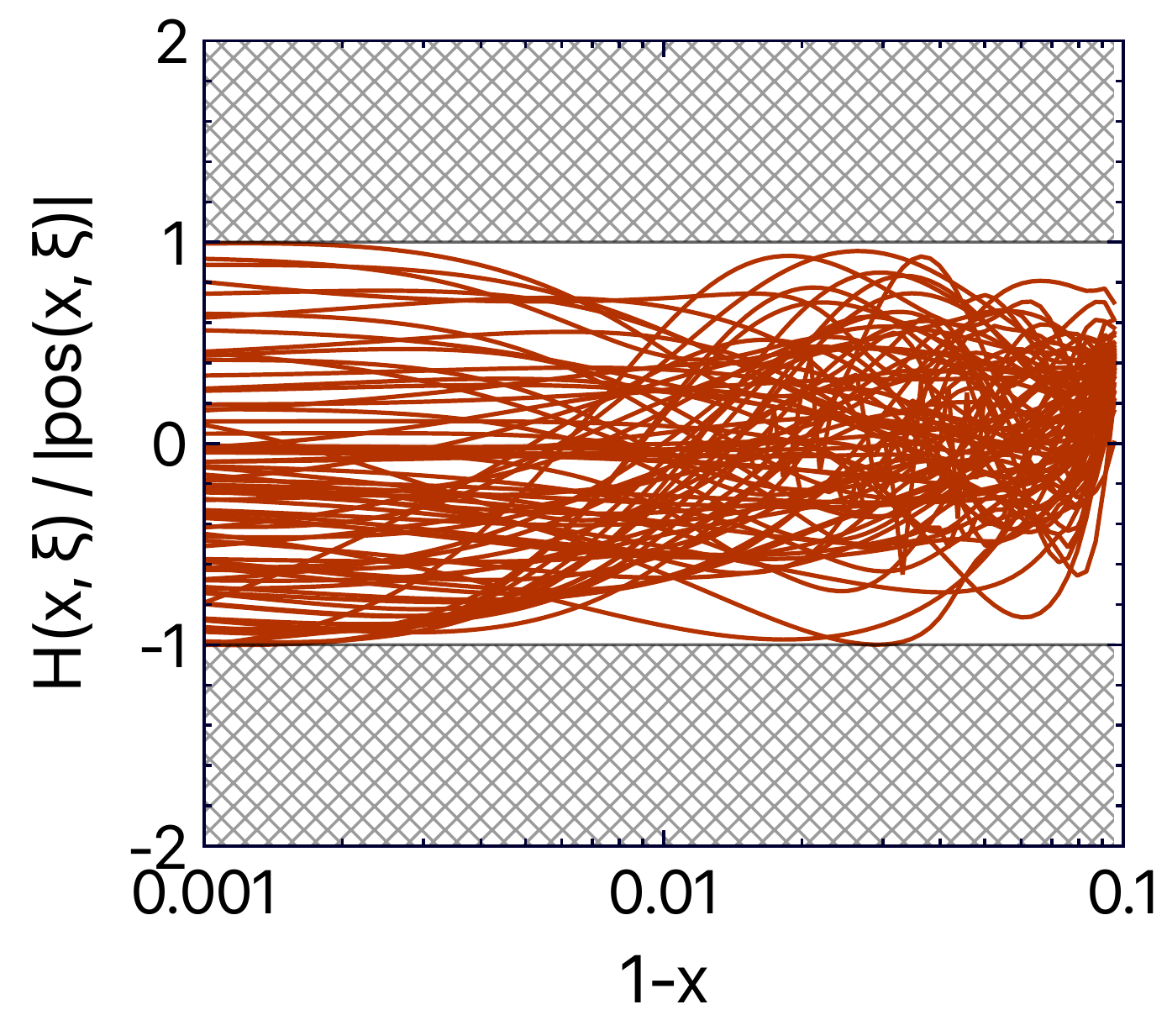}
\includegraphics[width=0.35\textwidth]{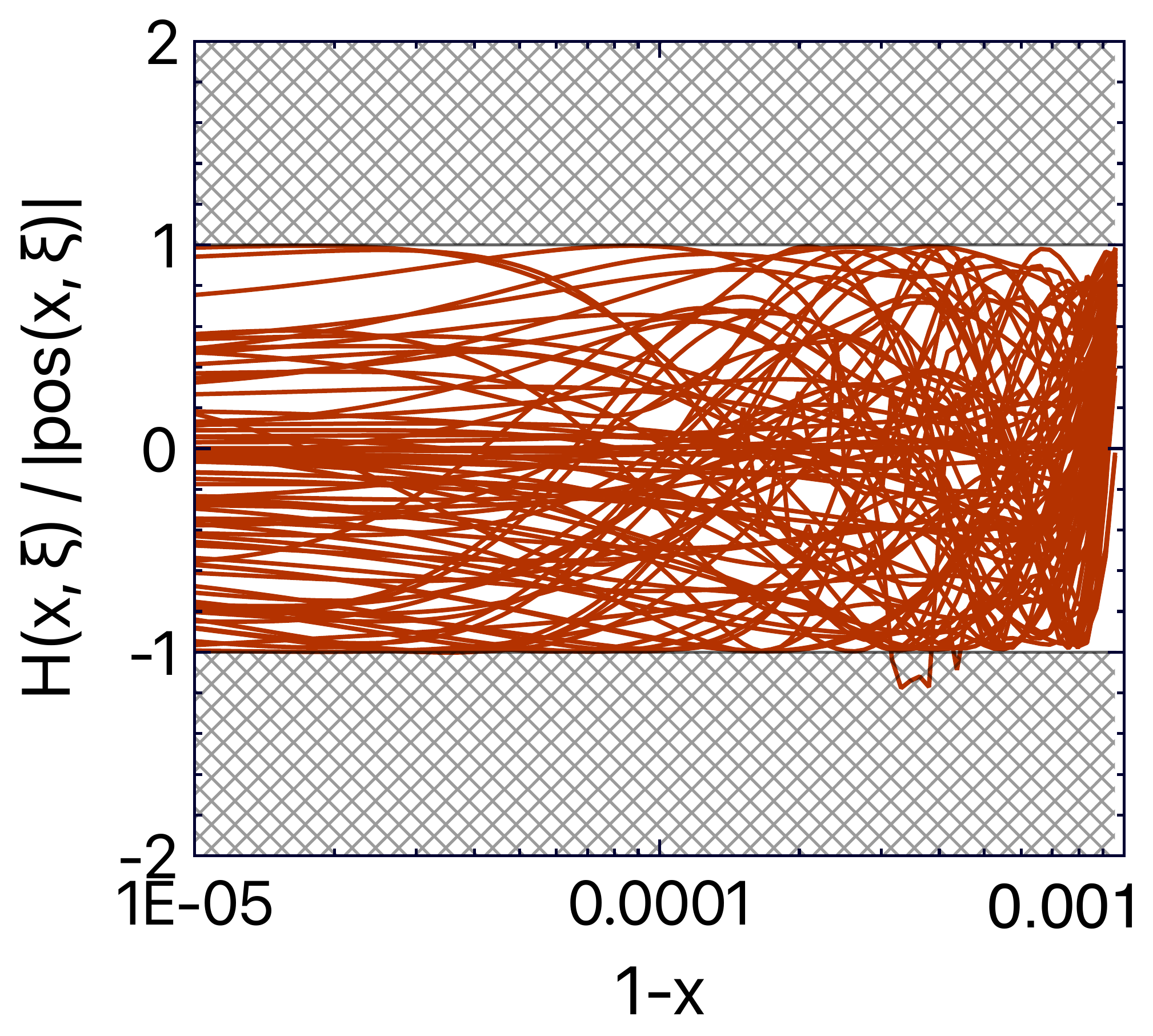}
\caption{Result of constraining the ANN model using $200$ points evaluated with the GK model \cite{Goloskokov:2005sd,Goloskokov:2007nt,Goloskokov:2009ia} for sea quarks: $x = \xi$ case, positivity enforced. Distributions of GPDs normalised by the positivity bound are shown as a function of either  $x$ or $(1-x)$. Solid red curves are for replicas used in this analysis. The regions excluded by the positivity constraint are denoted by the hatched bands. The plots are for (top left) $\xi = 0.0001$, (top right) $\xi = 0.01$, (bottom left) $\xi = 0.9$ and (bottom right) $\xi = 0.999$}
\label{xEqxiPosPositivity}
\end{center}
\end{figure*}

The genetic algorithm allows for a straightforward implementation of penalisation methods to avoid unwanted results, in particular those not fulfilling binary-like conditions, like inequalities. We use this feature to enforce the positivity condition. To achieve this, for each set of free parameters (\ie for each candidate) we check by how much we can scale $F_{S}(\beta, \alpha)$ by changing the $N_{S}$ parameter to saturate the condition given by Eq.~\eqref{eq:theory:positivity}. We check this in $10000$ points covering the $(x, \xi < x)$ domain. If we are not able to scale $F_{S}(\beta, \alpha)$ to respect the constraint in all of those points, the candidate is discarded, \ie it is dropped from the population of candidates. If we can, we scale $F_{S}(\beta, \alpha)$ accordingly, \ie we take the highest allowed value of $|N_{S}|$. This simple method of enforcing the positivity constraints requires a substantial computing power, but gives satisfactory results. We conclude that up to the numerical noise the constraint is fulfilled, which we demonstrate with Fig.~\ref{xEqxiPosPositivity} showing the $x$-dependencies of replicas with respect to the positivity bound for few values of $\xi$. We note that after the minimisation, $F_{C}(\beta, \alpha)$ and $F_{S}(\beta, \alpha)$ may not fulfill alone the positivity constraint given by the chosen PDF, but they do in sum. It indicates that the inclusion of $F_{S}(\beta, \alpha)$ gives an extra freedom to achieve the positivity.

%% file: sec_summary.tex
\section{Summary}
\label{sec:summary}

The lack of GPD flexible parameterisations fulfilling all required theoretical constraints has been an obstacle slowing down the completion of global fits to experimental data comparable to those currently achieved in the PDF community.
Often GPD models are either too rigid to accommodate the measurements, or insufficiently constrained from the theoretical point of view.
Since the different theoretical requirements are hardly met in common analytic expressions, neural networks are appealing tools to obtain flexible, yet complex, parameterisations. 
To the best of our knowledge, our present study is the first providing concrete elements to actually build GPD models based on neural network techniques.

In this Article we discuss different strategies to model GPDs in $(x,\xi)$ and $(\beta, \alpha)$ spaces. 
In all cases GPD Mellin moments and polynomiality play a central role. 
We show in particular that a neural network description of double distributions is an efficient way to implement polynomiality, positivity, discrete symmetries, as well as the reduction to any freely chosen PDF in the forward limit. 

An increased control of systematic effects is highly desirable in front of the precision era of GPD physics and extractions. To this aim we also address questions of a practical nature when dealing with many parameters: existence of local minima in optimisation routines, presence of outliers in statistical data analysis, \etc
 
In view of future GPD fits to experimental DVCS data, we pay a special attention to singlet GPDs, and to contributions with both vanishing forward limit and vanishing DVCS amplitude at leading order in perturbation theory.
Our parameterisations may thus be used as tools to study the model dependency affecting extractions of GPDs and derived quantities, like GPD Mellin moments and the total angular momentum of hadrons.
Since these parameterisations fulfil theory driven constraints, they may be conveniently used in current and future analyses of GPDs, or in connection \eg to lattice QCD, either in $x$-space or through Mellin moments. 

%% file: sec_ack.tex
\begin{acknowledgements}
The authors thank Valerio Bertone, Feliciano Carlos De Soto Borrero, Jose Manuel Morgado Chávez, Markus Diehl, Cedric Mezrag, Pepe Rodriguez-Quintero, Jorge Segovia and Jakub Wagner for valuable discussions. H.M. acknowledges the support by the Polish National Science Centre with the grant no. 2017/26/M/ST2/01074. P.S. is supported by the Polish National Science Centre with the grant no. 2019/35/D/ST2/00272. The computing resources of {\'S}wierk Computing Centre, Poland are greatly acknowledged. This project was supported by the European Union's Horizon 2020 research and innovation programme under grant agreement No 824093. This work is supported in part in the framework of the GLUODYNAMICS project funded by the "P2IO LabEx (ANR-10-LABX-0038)" in the framework "Investissements d’Avenir" (ANR-11-IDEX-0003-01) managed by the Agence Nationale de la Recherche (ANR), France. 
\end{acknowledgements}